\documentclass[fleqn,usenatbib]{mnras}

\usepackage[T1]{fontenc}
\usepackage{ae,aecompl}

\usepackage{graphicx}
\usepackage{txfonts}
\usepackage{longtable}
\usepackage{rotating}
\usepackage{multirow} 
\usepackage{appendix}
\usepackage{latexsym}
\usepackage{lscape}
\usepackage{afterpage}
\usepackage{caption}
\usepackage{hyperref}
\usepackage{array}
\usepackage{float}
\usepackage{bookmark}

\title[NGC\,7469 as seen by MEGARA]{NGC\,7469 as seen by MEGARA: new results from high-resolution IFU spectroscopy}

\author[S.~Cazzoli et al.]{S.~Cazzoli$^{1}$\thanks{E-mail: sara@iaa.es}, A.~Gil de Paz$^{2,3}$,  I.~M{\'a}rquez$^{1}$, J.~Masegosa$^{1}$, J.~Iglesias$^{1}$, J.~Gallego$^{2,3}$, \newauthor  E.~Carrasco$^{4}$, R.~Cedazo$^{5}$, M.L.~Garc{\'i}a-Vargas$^{6}$, {\'A}.~Castillo-Morales$^{2,3}$, S.~Pascual$^{2,3}$,  \newauthor N.~Cardiel$^{2,3}$, A.~P{\'e}rez-Calpena$^{6}$, P.~G{\'o}mez-Alvarez$^{6}$, I.~Mart{\'i}nez-Delgado$^{6}$ and \newauthor  L.~Hermosa-Mu\~{n}oz$^{1}$ \\
$^{1}$ IAA - Instituto de Astrof{\'i}sica de Andaluc{\'i}a (CSIC), Apdo. 3004, 18008, Granada, Spain\\
$^{2}$ Departamento de F{\'i}sica de la Tierra y Astrof{\'i}sica, Universidad Complutense de Madrid, E-28040 Madrid, Spain\\ %
$^{3}$ Instituto de F{\'i}sica de Part{\'i}culas y del Cosmos IPARCOS, Facultad de Ciencias F{\'i}sicas, Universidad Complutense de Madrid, E-28040 Madrid, Spain \\
$^{4}$ INAOE - Instituto Nacional de Astrof{\'i}sica, {\'O}ptica y Electr{\'o}nica, Luis Enrique Erro No.1, C.P. 72840, Tonantzintla, Puebla, Mexico\\
$^{5}$ Universidad Polit{\'e}cnica de Madrid, Madrid, Spain \\
$^{6}$ Fractal, S.L.N.E., Madrid, Spain\\
}
\date{Accepted 2020 February 07. Received in original form 2019 September 12}

\pubyear{2020}

\begin{document}
\label{firstpage}
\pagerange{\pageref{firstpage}--\pageref{lastpage}}
\maketitle
\begin{abstract}
We present our analysis of high-resolution (R\,$\sim$\,20\,000) GTC/MEGARA integral-field unit spectroscopic observations, obtained during the commissioning run, in the inner region (12.5\,arcsec\,$\times$\,11.3\,arcsec) of the active galaxy NGC\,7469, at spatial scales of 0.62\,arcsec. We explore the kinematics, dynamics, ionisation mechanisms and oxygen abundances of the ionised gas, by modelling the H$\alpha$-[N\,{\small II}] emission lines at high signal-to-noise ($>$\,15) with multiple Gaussian components. MEGARA observations reveal, for the first time for NGC\,7469, the presence of a very thin (20\,pc) ionised gas disc supported by rotation (V/$\sigma$\,=\,4.3), embedded in a thicker (222\,pc), dynamically hotter  (V/$\sigma$\,=\,1.3) one. These discs nearly co-rotate with similar peak-to-peak  velocities (163\,vs.\,137\,km\,s$^{-1}$), but with different average velocity dispersion (38\,$\pm$\,1 vs. 108\,$\pm$\,4\,km\,s$^{-1}$). The kinematics of both discs could be possibly perturbed  by star-forming regions. We interpret the morphology  and the kinematics of a third (broader) component ($\sigma$\,$>$\,250\,km\,s$^{-1}$) as suggestive of the presence of non-rotational turbulent motions possibly associated either to an outflow or to the lense. For the narrow component, the [N\,{\small II}]/H$\alpha$ ratios point to the  star-formation as the dominant mechanism of ionisation, being consistent with ionisation from shocks in the case of the intermediate component. All components have roughly solar metallicity. In the  nuclear region of NGC\,7469, at \textit{r}\,$\leq$\,1.85\,arcsec,  a very broad (FWHM\,=\,2590\,km\,s$^{-1}$)  H$\alpha$ component is contributing (41 per cent) to the global H$\alpha$-[N\,{\small II}] profile,  being originated in the (unresolved) broad line region of the Seyfert\,1.5 nucleus of  NGC\,7469.   

\end{abstract}



\section{Introduction}

\begin{figure*}
\includegraphics[width=.995\textwidth]{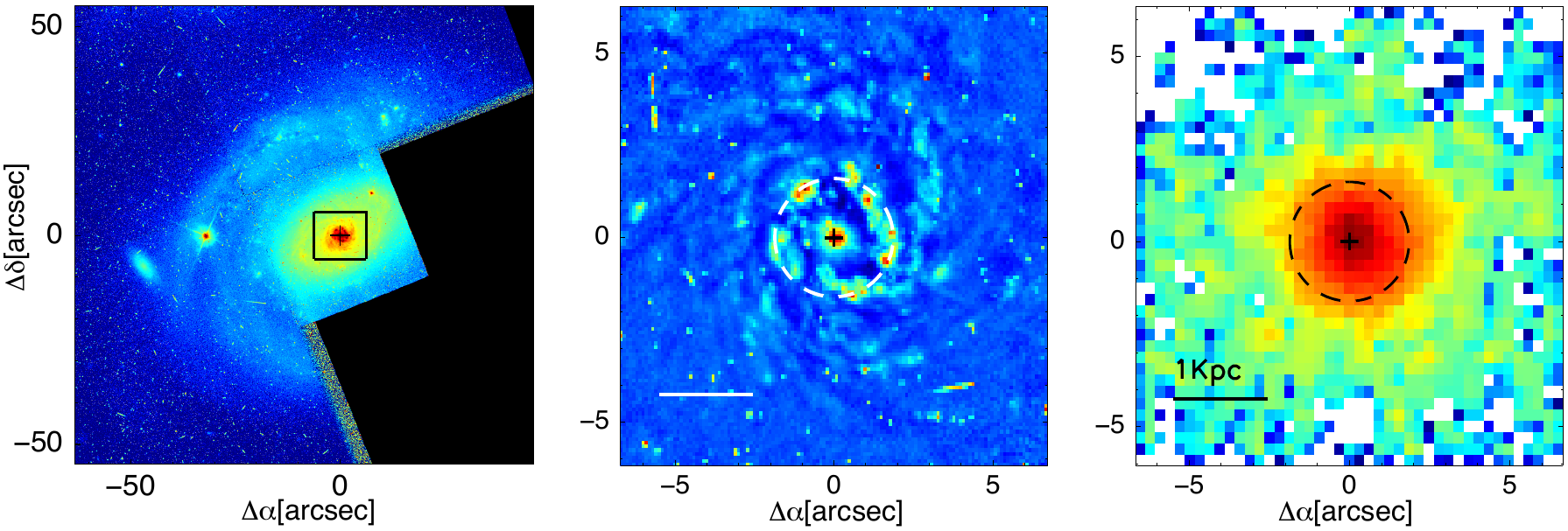}\\
\caption{Optical images of NGC\,7469. The left panel shows the large scale \textit{HST}-WFPC2 image (F606W filter), whereas the middle panel displays  its sharp-divided image (Sect.\,\ref{observations_datared}) with a zoomed-in view.  The zoomed image matches the field of view of MEGARA observations marked with a black square in the left panel. In the right panel, the optical  continuum from MEGARA IFU datacube (Sect.\,\ref{observations_datared}).  In all panels, the cross marks the galaxy nucleus (i.e. the peak of the continuum emission, see also Sect.\,\ref{observations_datared}). Horizontal bars show a linear scale of 1\,kpc (see Table\,\ref{T_properties}). The dashed circles indicates the nuclear region (Sect.\,\ref{observations_datared}).}
\label{HST_morphology} 	
\end{figure*}

Long slit and multi-object spectroscopy have been traditionally used to obtain spectra for both point and extended sources. While both techniques are efficient for single and unresolved sources, spatially resolved spectroscopy is needed for an optimal study of extended resolved sources. Integral Field Spectroscopy (IFS) has the advantage of being able to provide a spectrum of a spatial element in a two-dimensional field over a potentially long wavelength coverage.  Nowadays, the number of Integral Field Units (IFUs) is growing and many detailed studies of individual objects and large IFS surveys are being carried out (e.g. CALIFA, \citealt{Sanchez2012}; SAMI, \citealt{Croom2012}; MaNGA, \citealt{Bundy2015}; SINS, \citealt{Forster2009}; KMOS-3D,\citealt{Wisnioski2015}). However, all these studies are limited by the spectral resolution (generally less than 6000). This limits the detection and the kinematic analysis of galaxy component that might be detected as rather narrow but blended components. This is especially relevant for the H$\alpha$-[N\,{\small II}] complex, the most studied optical tracer, due to the proximity of the lines (15-20\,\AA). With the advent of MEGARA (\textit{Multi-Espectr{\'o}grafo en GTC de Alta Resoluci{\'o}n para Astronom{\'i}a}, \citealt{GilDePaz2014, GilDePaz2018, Carrasco2018}) the new integral field unit (IFU) at the 10.4m Gran Telescopio Canarias (GTC) high spectral resolution (up to 20\,000)  IFS observations have become available. MEGARA started the operation in 2017 and during the commissioning run the NGC\,7469
galaxy has been observed among others (e.g. NGC\,7025  \citealt{Dullo2019} and UGC\,10205  \citealt{CatalanTorrecilla2019}).\\
\noindent  NGC\,7469 (IRAS\,23007+0836, MRK\,1514) is a well studied  active galaxy in the local Universe (\textit{z}\,=\,0.016317). Table\,\ref{T_properties} summarizes the basic properties for this object and in Fig.\,\ref{HST_morphology} its optical morphology is shown from \textit{Hubble Space Telescope}  (\textit{HST}) observations. \\
\noindent  The Seyfert\,1.5 nucleus of  NGC\,7469 \citep{Landt2008} is hosted by a spiral galaxy (Fig.\,\ref{HST_morphology}). NGC\,7469 forms part of a galaxy pair together with IC\,5283 at a projected distance of 60\,-\,70\,Mpc \citep{Marquez1994}. Their past interaction, occurred 1.5\,$\times$\,10$^{8}$ year ago \citep{Genzel1995}, is  believed to have triggered the powerful star formation activity of NGC\,7469  (SFR\,=\,48\,M$_{\sun}$/yr, \citealt{PereiraSantaella2011}), mainly occurring in the circumnuclear star-forming ring bright at various wavelengths: radio (e.g. \citealt{Colina2001, Davies2004}), optical (e.g. \citealt{Scoville2000,  Colina2007}) and infrared (e.g. \citealt{Soifer2003, DiazSantos2007}). Such prominent star formation activity along with an high IR-luminosity (log\,(L$_{\rm\,IR}$/L$_{\odot}$)\,=\,11.7, \citealt{Sanders2003}) led NGC\,7469 to be enclosed among Luminous Infrared Galaxies (LIRGs).\\
\noindent  The spatially resolved disc kinematics has been studied mainly through ALMA and VLT/SINFONI observations of cold and warm molecular gas and ionised gas by  different authors \citep{Davies2004, Hicks2009, AlonsoHerrero2009, Fathi2015, Izumi2015}. Briefly these works found that the velocity fields obtained from the modelling of HCN, CO, H$_{2}$$\lambda$2.12$\mu$m, H$\alpha$$\lambda$6563 and Br$\gamma$$\lambda$2.165$\mu$m emission lines are suggestive of uniform large-scale rotation stable against star formation,  with no indication of warps or bar. However, a lense is found to be relevant at distances between 5.0\,arcsec and  15.0\,arcsec from the nucleus \citep{Marquez1994}.\\
\noindent   Non rotational motions, such as outflows are thought to be common in galaxies (\citealt{Veilleux2005} for a review). Growing evidence of outflows has been collected in the Milky  Way  (e.g. \citealt{Bordoloi2017, Fox2015, Fox2019}) and external galaxies, i.e. starbursts (e.g. \citealt{Heckman2000, Chen2010, Westmoquette2011}), LIRGs (e.g. \citealt{Arribas2014, Cazzoli2014, Cazzoli2016, PereiraSantaella2017}) and AGNs (e.g. \citealt{Harrison2014, Fiore2017, Maiolino2017, HG2018, Cazzoli2018}) at any redshift. In the case of NGC\,7469, \citet{MullerSanchez2011} found a wide angle nuclear outflow (half opening angle of 45$^{\circ}$) with a (maximum) velocities $<$\,200\,km\,s$^{-1}$  by analysing [S\,{\small IV}] coronal line at 1.96\,$\mu$m in the central 0.8\,arcsec\,$\times$\,0.8\,arcsec with SINFONI/VLT data. Evidence of the presence of a high velocity  outflow (V\,$\geq$\,580\,km\,s$^{-1}$) are found at X-rays (\cite{Blustin2007} and reference therein).\\ 

\noindent  In this paper, we used optical high spectral resolution (R\,$\sim$\,20\,000) MEGARA/GTC IFS observations to investigate the presence of the multiple distinct kinematic components of the H$\alpha$-[N\,{\small II}]$\lambda$$\lambda$6548,6584 complex associated to disc kinematics or non-rotational motions (e.g. outflows and/or chaotic motions associated to the lense) in NGC\,7469. Our main goals are to characterize all the components by studying their kinematics and dynamics, as well as exploring their possible dominant ionisation mechanism and infer oxygen abundances. Furthermore, we aim at investigating the properties of  H$\alpha$ broad  component originated in the broad line region  (BLR) of the AGN of NGC\,7469.\\

\noindent This paper is organized as follows. In Section  \ref{observations_datared}, the data and observations are presented as well as the data reduction. In Section\,\ref{emlines_maps}, we present the spectroscopic analysis: line modelling and maps  generations. The details of the kinematic analysis are given in Section\,\ref{Kinematic_Analysis}.  In Section\,\ref{main_observational_results},  the main observational results are highlighted. In Section\,\ref{discussion}, we discuss the origin, kinematics, dynamics,  the ionisation mechanisms and oxygen abundances of the different components considered in the line modelling. Finally, the main conclusions are presented in Section \ref{conclusions}.\\

\noindent  Throughout the paper we will assume H$_{\rm\,0}$\,=\,70\,km\,s$^{-1}$Mpc$^{-1}$ and the standard $\Omega_{\rm\,m}$\,=\,0.3,  $\Omega_{\rm\,\Lambda}$\,=\,0.7 cosmology. \\
All images and spectral maps are oriented following the standard criterium so the north is up and east to the left.
 
\begin{table}
\caption{General properties of NGC\,7469.} 
\begin{tabular}{l c c }
\hline
Properties    & Value & References \\
\hline
R.A. & 23$^{\rm h}$03$^{\rm m}$15$^{\rm s}$.6 & \\
Decl. & +08$^{\rm d}$52$^{\rm m}$26$^{\rm s}$&  \\
\textit{z} & 0.016317 &  NED\\
Distance [Mpc] & 70.2 & NED\\
Scale [pc/arcsec] & 340 & NED\\
Hubble classification &(R')SAB(rs)a & \citealt{Galbany2016}\\
Activity & Sy1.5 & \citealt{Landt2008}\\
Log\,(L$_{\rm\,IR}$/L$_{\odot}$) & 11.7 & \citealt{Sanders2003}\\
SFR[M$_{\sun}$/yr]& 48 & \citealt{PereiraSantaella2011}\\
\textit{i} [$^{\circ}$]& 30.2 & Hyperleda\\
PA$_{\rm\,phot}$ [$^{\circ}$]& 126& Hyperleda\\
\hline\\
\end{tabular}
\label{T_properties}

{\textit{Notes.}  \lq Distance\rq \ and \lq scale\rq \ are from the Local Group, as listed in the NASA Extragalactic Database, NED\footnote{\url{https://ned.ipac.caltech.edu/}}.  \lq L$_{\rm\,IR}$\rq \ corresponds to the L$_{\rm\,8-1000}$ and it is calculated using the fluxes in the four IRAS bands as given in \citet{Sanders2003}. \lq SFR\rq \ is the  star-formation rate based on the 24\,$\mu$m luminosity. The AGN contribution to the 24\,$\mu$m  luminosity is subtracted. The inclination  position angle  and \lq \textit{i}\rq \ and \lq PA$_{\rm\,phot}$\rq ,  are defined as the inclination between line of sight and polar axis of the galaxy and the major axis position angle (north eastwards), respectively.More specifically,  \lq PA$_{\rm\,phot}$\rq \ is the average value between the measurements from \citet{Paturel2000} and the \lq Uppsala General Catalogue of Galaxies\rq \ \citep{Nilson1973} and \textit{i} is determined from the axis ratio of the isophote in the B-band  using  a correction for intrinsic  thickness based on the morphological type (see references in Hyperleda for individual measurements).} 
\end{table}


\footnotetext{\url{https://ned.ipac.caltech.edu/}}

\section[]{Observations and data reduction}
\label{observations_datared}

Observations were carried out with GTC/MEGARA on 2017 July 25 during the MEGARA commissioning run. The 567  fibers that constitute the MEGARA IFU (100\,$\mu$m in core size) are arranged on a square microlens-array that projects on the sky a field of 12.5\,arcsec\,$\times$\,11.3\,arcsec in size. Each microlens is a hexagon inscribed in a circle  with  diameter of 0.62\,arcsec projected in the sky. A total of 56 ancillary  fibers (organized in eight fiber bundles), located  at a distance of 1.75-\,2.0\,arcmin from the center of the IFU field of view, deliver simultaneous sky observations.\\
 We used the  HR-R Volume-Phased Holographic (VPH, i.e. VPH665-HR) covering the 6405.6\,--\,6797.1\,\AA \ spectral range  with a spectral resolution of R\,$\sim$\,20\,000 (from 19\,500 to 20\,600). The linear dispersion is of $\sim$\,0.0974\,\AA/pixel , hence $\sim$\,5\,km\,s$^{-1}$ at the corresponding wavelength of H$\alpha$$\lambda$6563.  We obtained 3 exposures of 600\,s each to facilitate cosmic-rays removal.  During the observations of NGC\,7469 the dimm-seeing was 1.02\,arcsec; airmass was 1.28.\\
MEGARA raw  data were reduced with the data reduction package provided by Universidad Complutense de Madrid  (\textsc{megara drp}\footnote{\url{https://github.com/guaix-ucm/megaradrp/}}; version 0.8; \citealt{Pascual2018,Pascual2019})  following the MEGARA cookbook\footnote{\url{https://doi.org/10.5281/zenodo.1974954}}. The pipeline allows to perform the following steps: sky and bias subtraction, flat field correction, spectra tracing and extraction, correction of fiber and pixel transmission, wavelength calibration. We did not apply any flux calibration, as we are mainly interested in the kinematic analysis.   \\
We  apply a regularization grid to obtain square spaxels\footnote{Hereafter, as  \lq spaxels\rq, we will refer to all the spatial elements (pixel) in the cube obtained after the application of the regularization grid (not those from the fully reduced RSS-cube).} of size of 0.31\,arcsec; the final cube has dimensions of 43\,$\times$\,40\,$\times$\,4300, equivalent to a total of 1720 spectra in the datacube. \\
The point spread function (PSF) of the MEGARA datacube could be described as Moffat function \citep{Moffat1969} and it has been estimated to have a full width at half-maximum (FWHM) of 0.93\,arcsec from the 2D-profile brightness distribution of standard stars. Throughout the paper, in order to avoid any possible PSF contamination in the kinematic measurements, we will conservatively consider as \lq nuclear region\rq \ a circular area of radius equal to the width at 5 per cent intensity of the PSF radial profile, i.e. 1.85\,arcsec (see Fig.\,\ref{HST_morphology}). This area nearly coincides with the outer radius of the circumnuclear star forming ring and it is marked (with a dashed circle) in the images and spectral maps from the MEGARA cube, as well as, in the \textit{HST} image in the middle panel of Fig.\,\ref{HST_morphology}. \\
We checked the wavelength calibration and obtained the instrumental profile using two OH sky-lines ($\lambda$6499 and  $\lambda$6554) bluewards to  H$\alpha$-[N\,{\small II}]; sky-lines are absent in the red part of the spectra. The mean central wavelengths were found to be 6498.725\,$\pm$\,0.046\,\AA \ and 6553.621\,$\pm$\,0.047\,\AA \ coinciding well with their actual values i.e. 6498.729\,\AA \ and 6553.617\,\AA  \ from \citet{Osterbrock1996}. The width ($\sigma$) of the two sky-lines (instrumental dispersion) is uniform  over the entire field of view with average and standard deviation values of 0.12\,$\pm$\,0.05\,\AA \ and 0.14\,$\pm$\,0.05\,\AA.   
The results of these checks indicate negligible systematic calibration errors will affect our measurements of the ionised gas kinematics (Sect.\,\ref{emlines_maps}).
Hence, hereafter, the fitting errors (generally larger than 0.1\,\AA) would be considered as the main source of uncertainty.\\
For each spectrum (i.e. on spaxel-by-spaxel basis), the effect of instrumental dispersion ($\sigma_{\rm\,INS}$) was corrected for by subtracting it in quadrature from the observed line dispersion ($\sigma_{\rm\,obs}$) i.e. $\sigma_{\rm\,line}$\,=\,$\sqrt{\sigma_{\rm\,obs}^{2}\,-\,\sigma_{\rm\,INS}^{2}}$. \\
The field of view of MEGARA observations is overlaid to the \textit{HST}-WFPC2 (Wide-Field Planetary Camera 2) image shown in the left panel of Fig.\,\ref{HST_morphology}. In order to highlight features departing from axisymmetry, we produced the \lq sharp-divided\rq \ image (Fig.\,\ref{HST_morphology}, middle). It is obtained by dividing the original image, \textit{I}, by a filtered version of it  \citep{Marquez1999,Marquez2003}. The image is median filtered with the \textsc{iraf}\footnote{\url{http://iraf.noao.edu}} command \lq median\rq \ using a box of 20 pixels. \\
The right panel of Fig.\,\ref{HST_morphology},  shows the optical continuum (6380\,--\,6480\,\AA, rest frame wavelength range) image generated from the MEGARA-IFU datacube. Throughout the paper,  we consider the peak of the continuum emission as the photometric center (nucleus) of the galaxy with an accuracy of 0.47\,arcsec (half of the FWHM of the PSF). It is marked as a cross in all images and spectral maps obtained from the MEGARA cube. 

\section[]{Emission lines modelling and maps generation}
\label{emlines_maps}
All the spectra in the cube were visually inspected  in order to check and correct for possible data reduction artifacts (e.g. residual comics rays), as well as to look for the presence of line asymmetries and complex line profiles. These latter are present at nearly any distance from the nucleus. Moreover, most of the spectra in the unresolved nuclear region of NGC\,7469 (at \textit{r}\,$\leq$\,1.85\,arcsec, i.e. $\sim$\,630\,pc at the adopted distance, see Table\,\ref{T_properties}),   show broad H$\alpha$$\lambda$6563 emission line superposed with strong narrow lines, which is typical for a Seyfert\,1 nucleus.\\
We excluded from the fitting procedure those spectra with signal-to-noise per pixel (SNR) in H$\alpha$ lower than 15  and those with artifacts. These are the 57 per cent of the total number of spaxels.\\ 
This conservative limit in SNR will allow  us to maximize the exploitation of the unprecedent spectral resolution of  MEGARA data.\\
We did not apply any procedure for the subtraction of the underlying stellar light for MEGARA spectra as their wavelength coverage (Sect.\,\ref{observations_datared}) prevents an optimal stellar-continuum modelling.


\subsection{Line fitting}
\label{line_fitting}

\begin{figure}
\includegraphics[width=.49\textwidth]{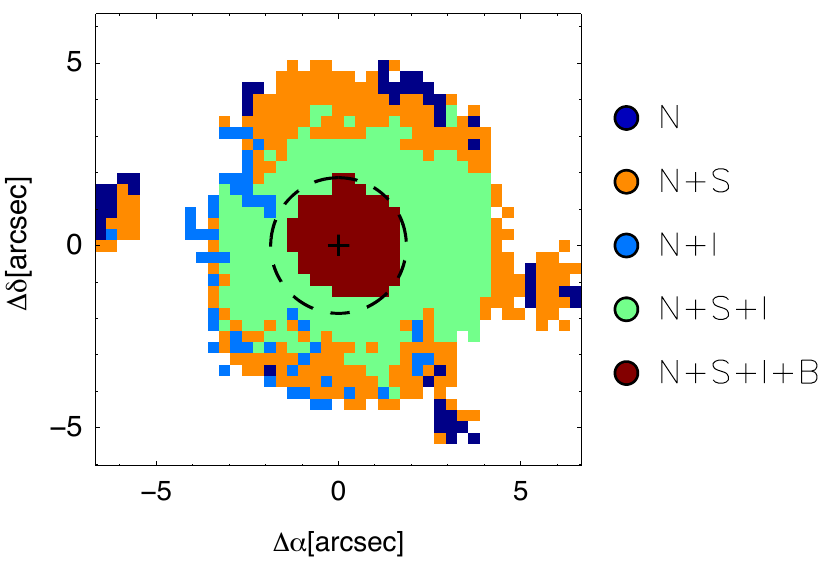}\\
\caption{Map of the number of components used to model the observed line profiles in the MEGARA cube (Sect.\,\ref{line_fitting}). The color coding  indicates the different combinations of components considered for the final fits. These are indicated in the legend where \lq N\rq ,\lq S\rq, \lq I\rq, and \lq B\rq \ stand for narrow, second, intermediate-width and broad components, respectively. For each of the five possible combinations of components, examples of the fits are shown in Fig.\,\ref{Fig_spaxels}.}   
\label{Fig_snr} 	
\end{figure}

\begin{figure}
   \includegraphics[trim = 5.cm 13.15cm 4.35cm 5.7cm, clip=true, width=.475\textwidth]{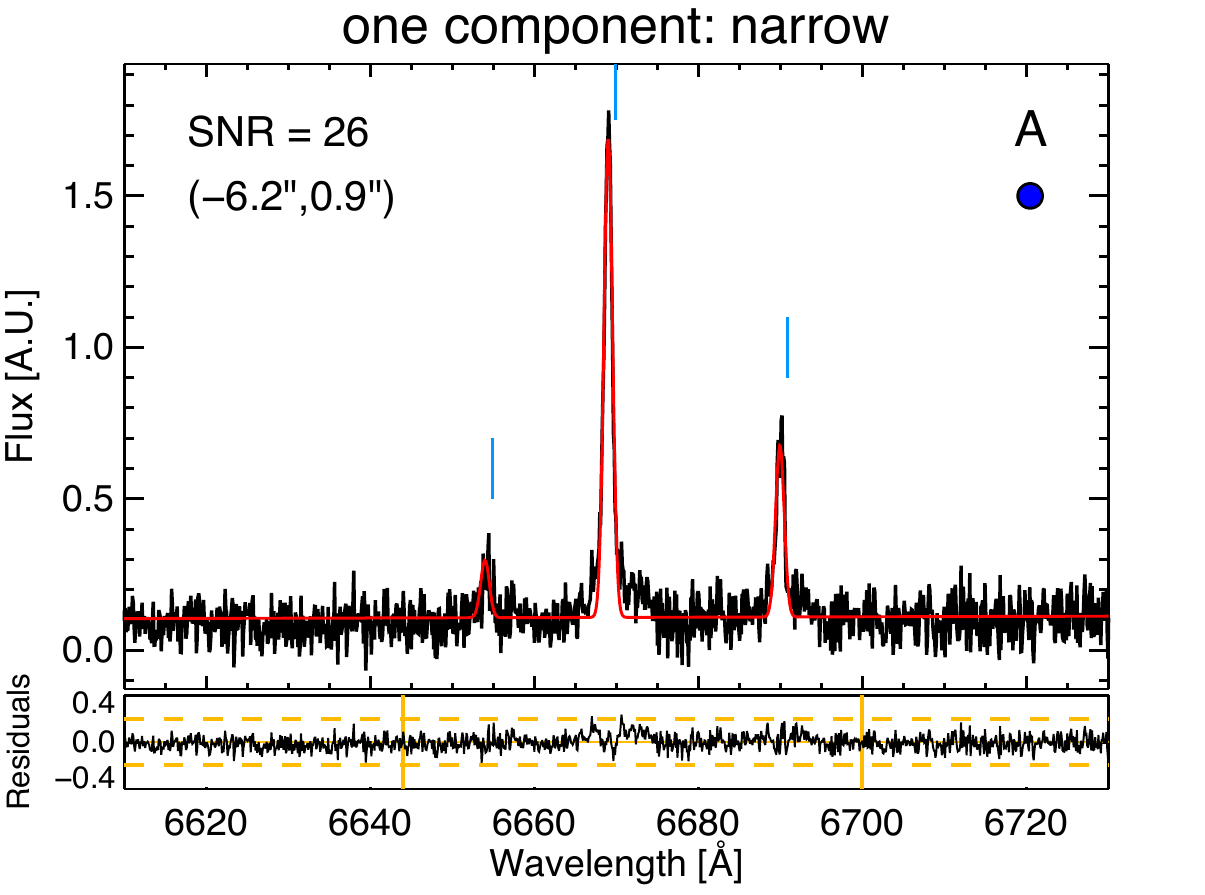}\\
\includegraphics[trim = 5.cm 13.15cm 4.35cm 5.7cm, clip=true, width=.475\textwidth]{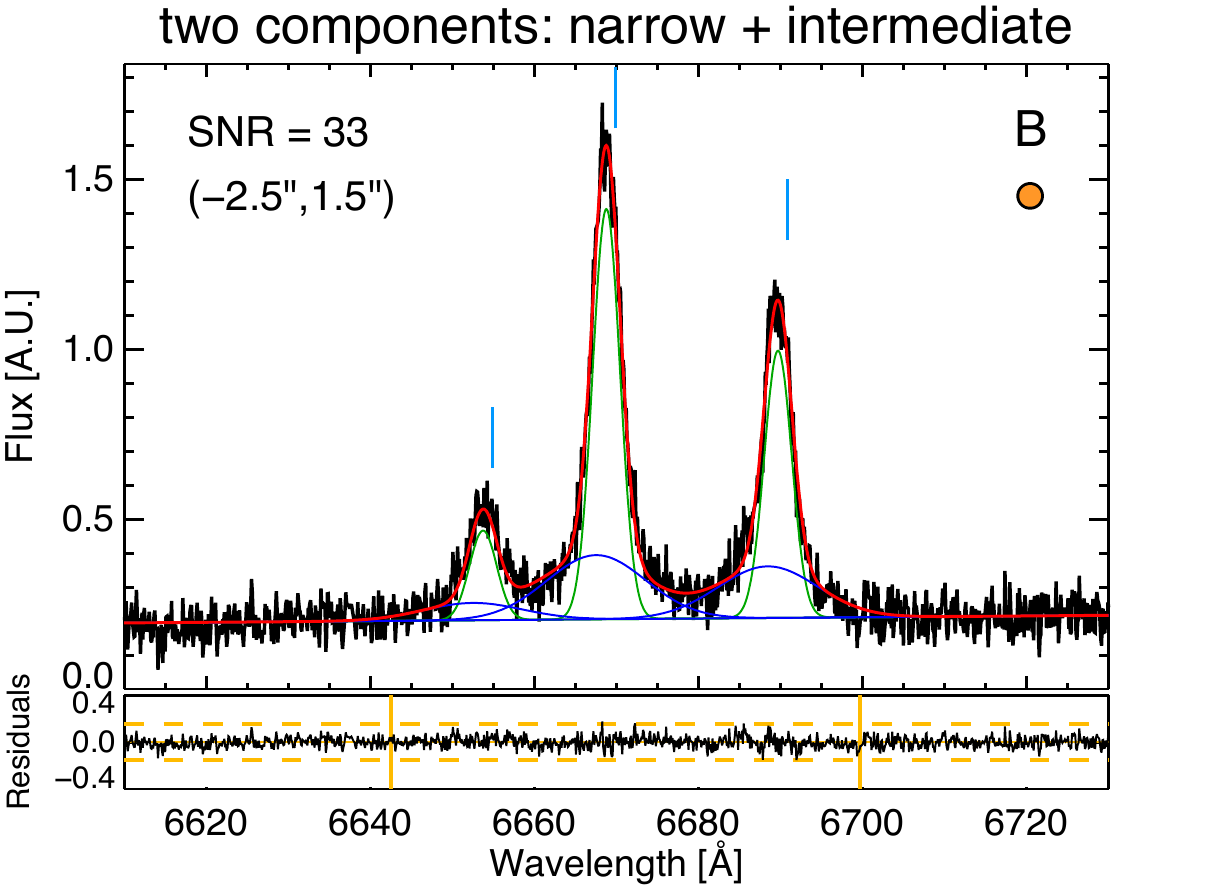}\\
\includegraphics[trim = 5.cm 13.25cm 4.35cm 5.7cm, clip=true, width=.475\textwidth]{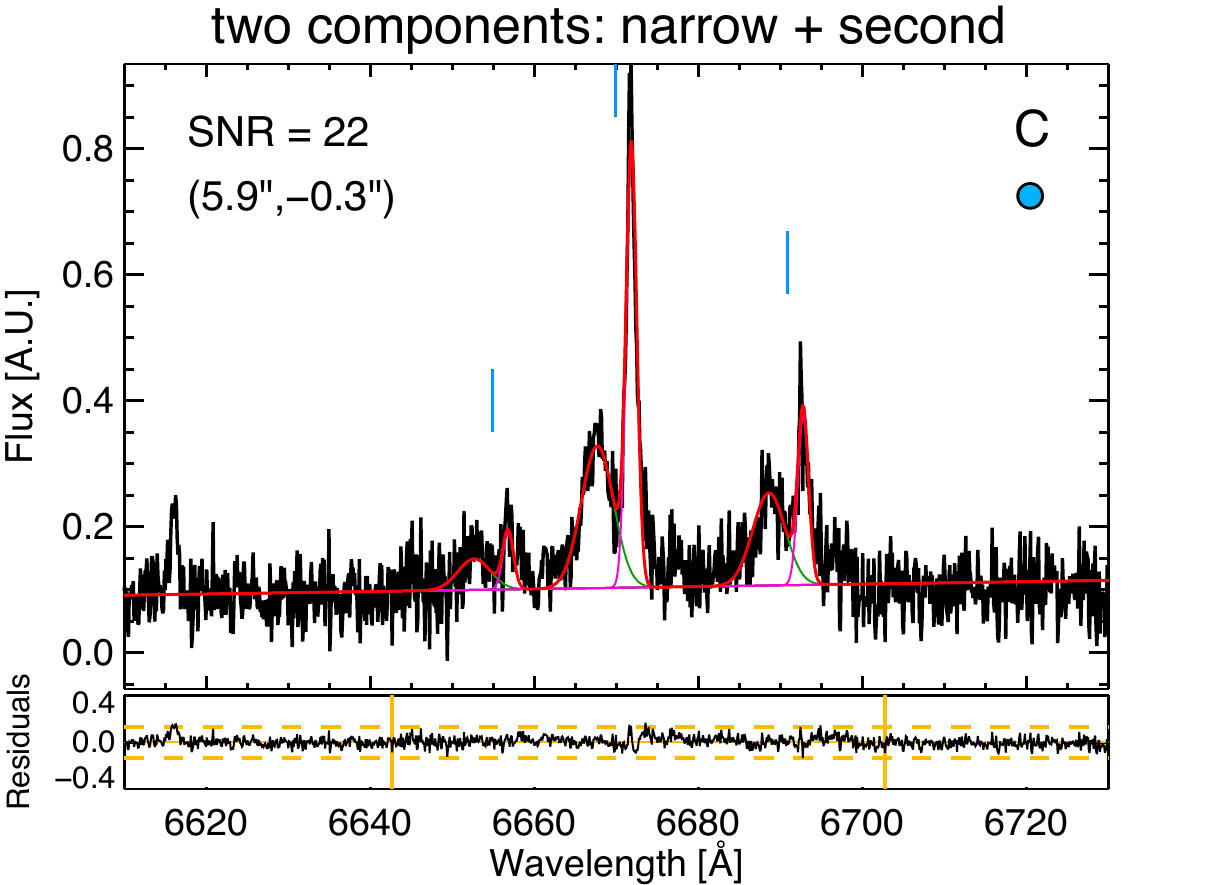}\\
\caption{Examples of H$\alpha$-[N\,{\small II}] observed spectra and their modelling for selected regions. The coordinate labels indicate distance from the photometric center (Sect.\,\ref{observations_datared}).  As reference, blue vertical lines mark  the systemic wavelengths of the emission lines. For each spaxel (from \lq A\rq \ to \lq E\rq \  panels) the modelled line profile (red continuous line) and the components (with different colors) are shown. Specifically, green, pink, blue Gaussian curves indicate: narrow, second and intermediate components used to model [N\,{\small II}] lines and narrow H$\alpha$; in purple is marked the broad H$\alpha$ component from the BLR (see Sect.\,\ref{line_fitting}). Same colours mark the same components, that are summarized at the top. Small circles follow the color coding of Fig.\,\ref{Fig_snr}.  Residuals from the fit (i.e. data\,$-$\,model) are in the small lower panels in which orange dashed lines indicate the $\pm$\,3$\times$$\varepsilon_{\rm\,c}$. Vertical orange continuos lines mark the wavelength range considered for calculating $\varepsilon_{\rm\,line}$ for the different cases (Sect.\,\ref{line_fitting}).}   
\label{Fig_spaxels} 	
\end{figure}

The  H$\alpha$ and  [N\,{\small II}]$\lambda$$\lambda$6548,6584 emission lines were simultaneously modelled with  Gaussian functions using a Levenberg-Marquardt least-squares fitting routine (\textsc{mpfitexpr})  by \citet{Markwardt2009},  within the Interactive Data Analysis\footnote{\url{http://www.harrisgeospatial.com/SoftwareTechnology/IDL.aspx}} (\textsc{idl}) environment. We tied all the lines to have  the same line-width and velocity shift. We   imposed also that the intensity ratio between the [N\,{\small II}]$\lambda$6548 and the [N\,{\small II}]$\lambda$6584 lines,  satisfies the 1:3 relation \citep{Osterbrock2006}.\\
The procedure for fitting H$\alpha$-[N\,{\small II}] takes into account possible multiple line components. Specifically, we realized that four distinct kinematic components (two narrows, one broad and one of intermediate width) were sufficient. This number of components ensures a trade off between the kinematical details, the SNR and the goodness of the fit (see \citealt{Bosch2019}). Specifically, four components represent the largest meaningful number of components to perform the kinematics analysis  throughout the galaxy.  Fig.\,\ref{Fig_snr} shows a map of the number/combinations of the components used during the line-fitting procedure. Figure\,\ref{Fig_spaxels} shows examples of the Gaussian fits. \\
The procedure is organized in different steps and is aimed at obtaining the best-fitting to the H$\alpha$-[N\,{\small II}] emission features. \\
To prevent overfit models and identify possible broad wings or double peaks, we cannot rely on the chi-square ($\chi$$^{^2}$) estimator as the goodness of the least-square minimization depends to the number of components \citep{Bosch2019}. Therefore, to allow for the appropriate number of Gaussians,  we followed the method proposed by \citet{Cazzoli2018} and already successfully applied to optical spectra of active galaxies (see also \citealt{Cazzoli2018}, \citealt{LHG2019} and \citealt{Hermosa2019}), on a spaxel-by-spaxel basis. More specifically, to assess whether the addition of a component is significant, we first calculated the standard deviation of a portion of the continuum (30\,\AA) free of emission lines, i.e. $\varepsilon_{\rm\,c}$. Then, throughout all the steps of the line fitting procedure (from one to four components),  we compared the $\varepsilon_{\rm\,c}$ value with the standard deviation estimated from the residuals under H$\alpha$-[N\,{\small II}], i.e. $\varepsilon_{\rm\,line}$, in a wavelength range\footnote{The wavelength ranges were selected in order to cover the H$\alpha$-[N\,{\small II}] complex and taking into account possible blue/red-shifted components.} of $\sim$\,60\,\AA \ ($\sim$\,100\,\AA \  if the broad component is included in the fitting). We considered a reliable fit with an adequate number of components when $\varepsilon_{\rm\,line}$\,$<$\,3\,$\times$\,$\varepsilon_{\rm\,c}$ (a \lq3-$\sigma$ significance\rq \ criterion), otherwise we tested the inclusion of an additional component.  Finally, we also checked  the results from the fits by visually inspecting all the residuals before and after the addition of a component. \\
The procedure is organized as follows. Initially, we fit all the emission lines to a single component, i.e. one Gaussian per each [N\,{\small II}]  line and narrow H$\alpha$ (hereafter, narrow component, Fig.\,\ref{Fig_spaxels}\,A).  In those cases where the observed line profiles show asymmetries (broad wings), the procedure adds  a second Gaussian component of intermediate width  (hereafter, intermediate component, Fig.\,\ref{Fig_spaxels}\,B). The two-Gaussian fits lead to two components which in general can be distinguished according to their widths i.e. narrow ($\sigma$\,$\leq$\,3\,\AA, mostly) and intermediate ($\sigma$\,$\geq$\,4.5\,\AA, generally) components. Then, we add a third H$\alpha$ component originated in the broad line region (BLR) of the AGN (hereafter, broad component), in most of the spectra within the nuclear region (see Sect.\,\ref{observations_datared}). 
This component can be clearly distinguished from the other two previously considered as the typical line width is larger than 17\,\AA \ (i.e. more than 40\,\AA \ in FWHM). Some of the spectra have  double-peaked and/or asymmetrical   profiles (e.g. Figures \ref{Fig_spaxels}\,C and \ref{Fig_spaxels}\,D) that suggest the presence of an additional narrow component (i.e. with a width lower than the intermediate-component). 
Therefore, for these spectra we used three  components (including the intermediate one) for the fit of [N\,{\small II}] lines and narrow H$\alpha$ (Fig.\,\ref{Fig_spaxels}\,E). In some cases, although the second component seemed necessary, the modelling of the H$\alpha$-[N\,{\small II}] lines was rather complicated leading  to residuals with substructures. Therefore, in those circumstances, in order to reproduce well the line profiles, the second component was constrained to have a width lower than 3.5\,\AA. These limits allow to constrain the overall procedure to deliver meaningful results lowering the residuals (measured in terms of $\varepsilon_{\rm\,FIT}$\,=\,$\varepsilon_{\rm\,line}$/$\varepsilon_{\rm\,c}$). \\
Overall, the final line modeling was satisfactory. The procedure reproduces well the observed line profiles with low residuals (see lower panels  in Fig.\,\ref{Fig_spaxels}).The few exceptions to this general behavior are seen  in the central region (i.e. within the region affected by the PSF marked as dashed circle in all maps, see Sect.\,\ref{observations_datared}), where although residuals are within the significance of $\pm$\,3$\times$$\varepsilon_{\rm\,c}$ some substructures seem to exist (e.g. Fig.\,\ref{Fig_spaxels}\,E). These features might indicate that a Gaussian-fit to the broad BLR-emission is a simplification (e.g. Lorentzian profiles, see \citealt{Kollatschny2013} and references therein). A detailed modelling of the (unresolved) BLR-emission is beyond the aim of the paper.\\
In general, the final values for  $\varepsilon_{\rm\,FIT}$  are never larger than 3, being larger than 2 only in  6 per cent of the cases (46 out of 738 total spaxels considered); on average $\varepsilon_{\rm\,FIT}$ is 1.2.\\
As a further check, in few representative example of the fit well outside the nuclear region (see Fig.\,\ref{Fig_spaxels}), we verified if the number of the selected Gaussian  components (N) is adequate, i.e. all statistically significant and there is no need of increasing the number of components. This test was done using the Akaike Information Criterion indicator (AIC, \citealt{Akaike1974}) which enable the comparison of models with different number of components (see \citealt{Bosch2019}). In all then cases, the number of components is very strong justified by a $\Delta$\,AIC defined as: AIC$_{\rm N+1}$\,-\,AIC$_{\rm  N}$, larger than 10 \citep{Wei2016}. \\
Formal fitting errors are the 1-sigma parameter uncertainties weighted with the square-root of the reduced $\chi$$^{^2}$ (see \textsc{mpfitexpr} documentation). These are larger than calibrations errors (see Sect.\,\ref{observations_datared} for the latter) on average, i.e. (0.2\,vs.\,0.12)\,\AA \ and (0.1\,vs.\,0.05)\,\AA \, for central wavelengths and widths respectively.

\begin{figure}
\includegraphics[trim = 5.cm 13.15cm 4.35cm 5.7cm, clip=true, width=.475\textwidth]{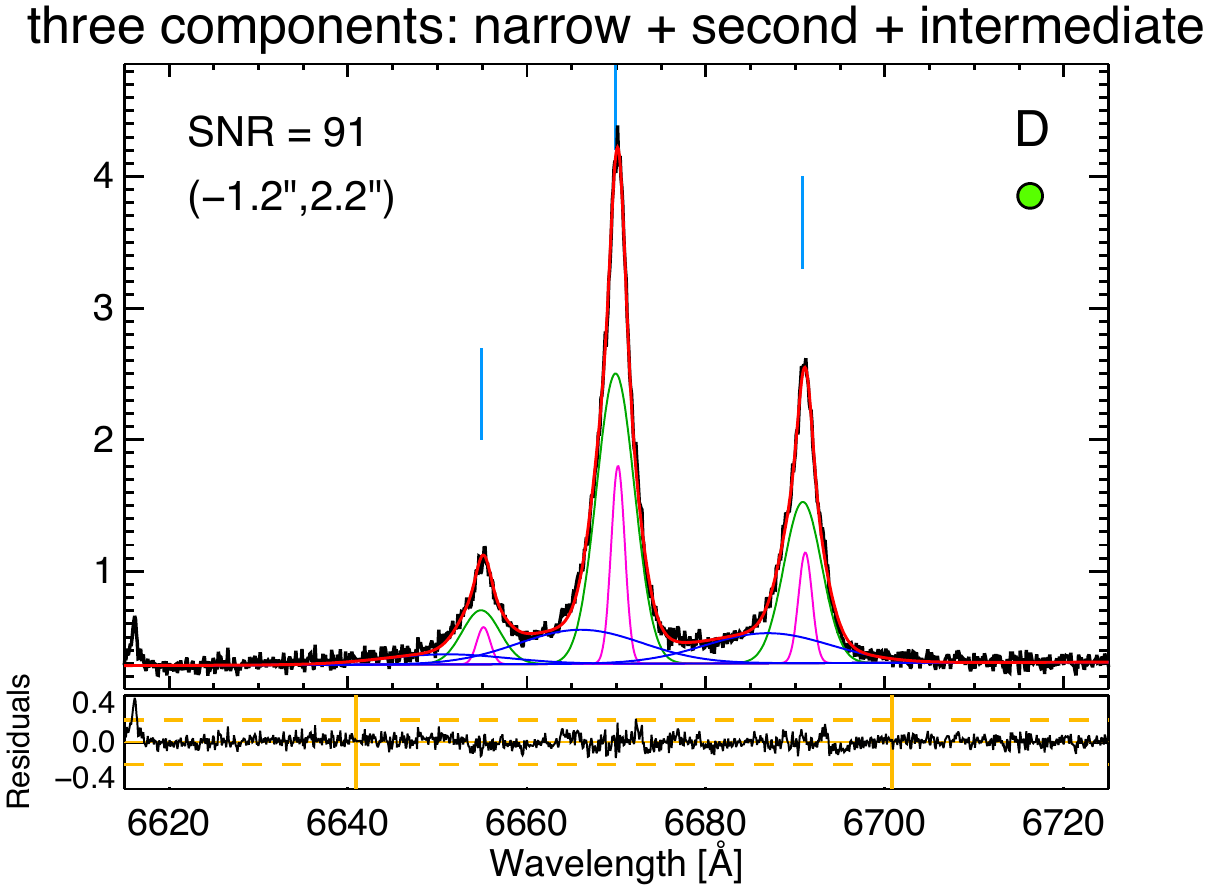}\\
\includegraphics[trim = 5.cm 13.25cm 4.35cm 5.7cm, clip=true, width=.475\textwidth]{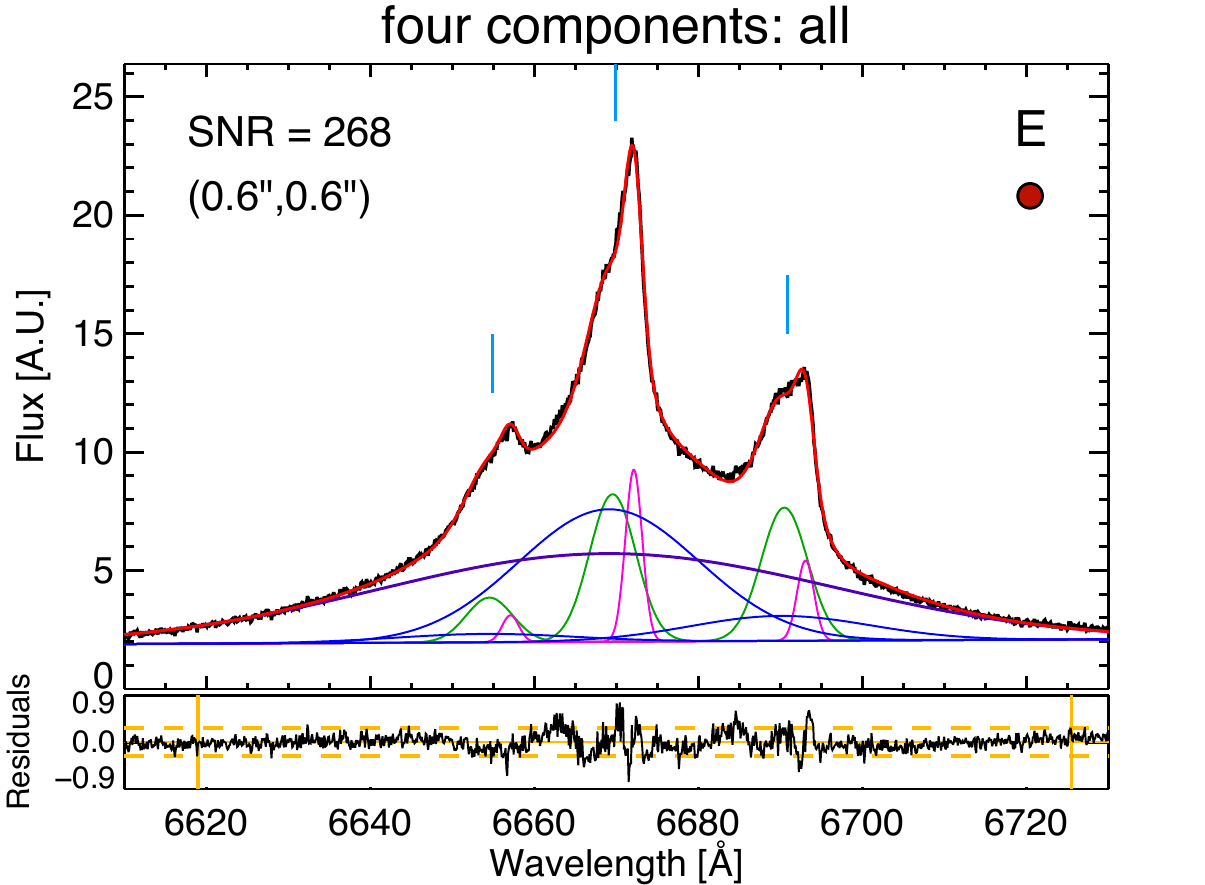}\\
 \captionsetup{labelformat=empty}{Fig.\,\ref{Fig_spaxels}\,--\,Continued.}   
\label{Fig_spaxels2} 	
\end{figure} 
\subsection{Generation of spectral maps}
\label{maps_generation}

From the MEGARA datacube, we generate the continuum image which is shown in Fig.\,\ref{HST_morphology} as  reference and in comparison to \textit{HST} images. \\
For each emission line and component  found with the procedure described in Sect.\,\ref{line_fitting}, we end up with the following information: central wavelength, width, and flux intensity along with their respective fitting errors. The intrinsic line widths were computed after removing the instrumental profile inferred from the sky lines (Sect.\,\ref{observations_datared}).\\

\begin{figure*}
\includegraphics[trim = 1.1cm .75cm 1.7cm 14.5cm, clip=true, width=1.\textwidth]{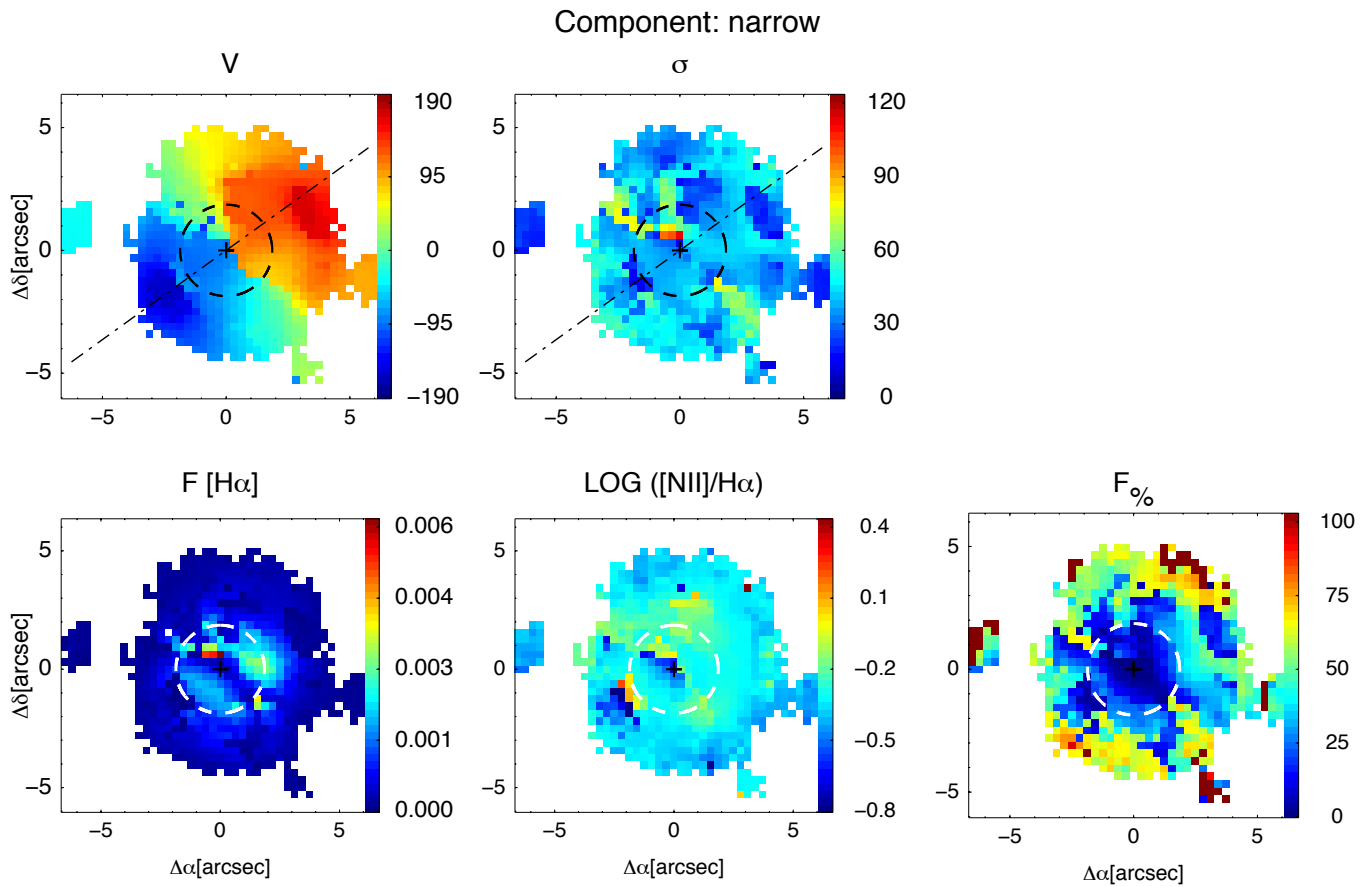}\\
\caption{Ionised gas maps  derived from the fit to the H$\alpha$-[N\,{\small II}] emission line profiles for the narrow component. \textit{Top panels}: kinematic maps, i.e. the velocity field (V) and the velocity dispersion map ($\sigma$). These are in km\,s$^{-1}$ units and they are color coded according to their own scale (i.e. range of the velocity and velocity dispersion) to facilitate the contrast and to highlight weak features. \textit{Bottom panels}: from left to right, H$\alpha$ flux-intensity map (F[H$\alpha$]), [N\,{\small II}]/H$\alpha$ line ratio and that of the percentage contribution of the narrow component to the total H$\alpha$-[N\,{\small II}] emission (F$_{\%}$). The former  is in arbitrary units as no flux-calibration has been applied (Sect.\,\ref{observations_datared}). The latter  is shown with the 0\,--\,100$\%$ range to enable the comparison with the same kind of map but for different components (Figures \ref{maps_2c} and \ref{maps_3c}  bottom right). The cross marks the photometric center,    the dashed  circle indicates the nuclear region (Sect.\,\ref{observations_datared}). The dot-dashed line is the major photometric axis oriented accordingly to photometric PA (Table\,\ref{T_properties}). }   
\label{maps_1c} 	
\end{figure*}

\begin{figure*}
\includegraphics[trim = 1.1cm .75cm 1.7cm 14.5cm, clip=true, width=1.\textwidth]{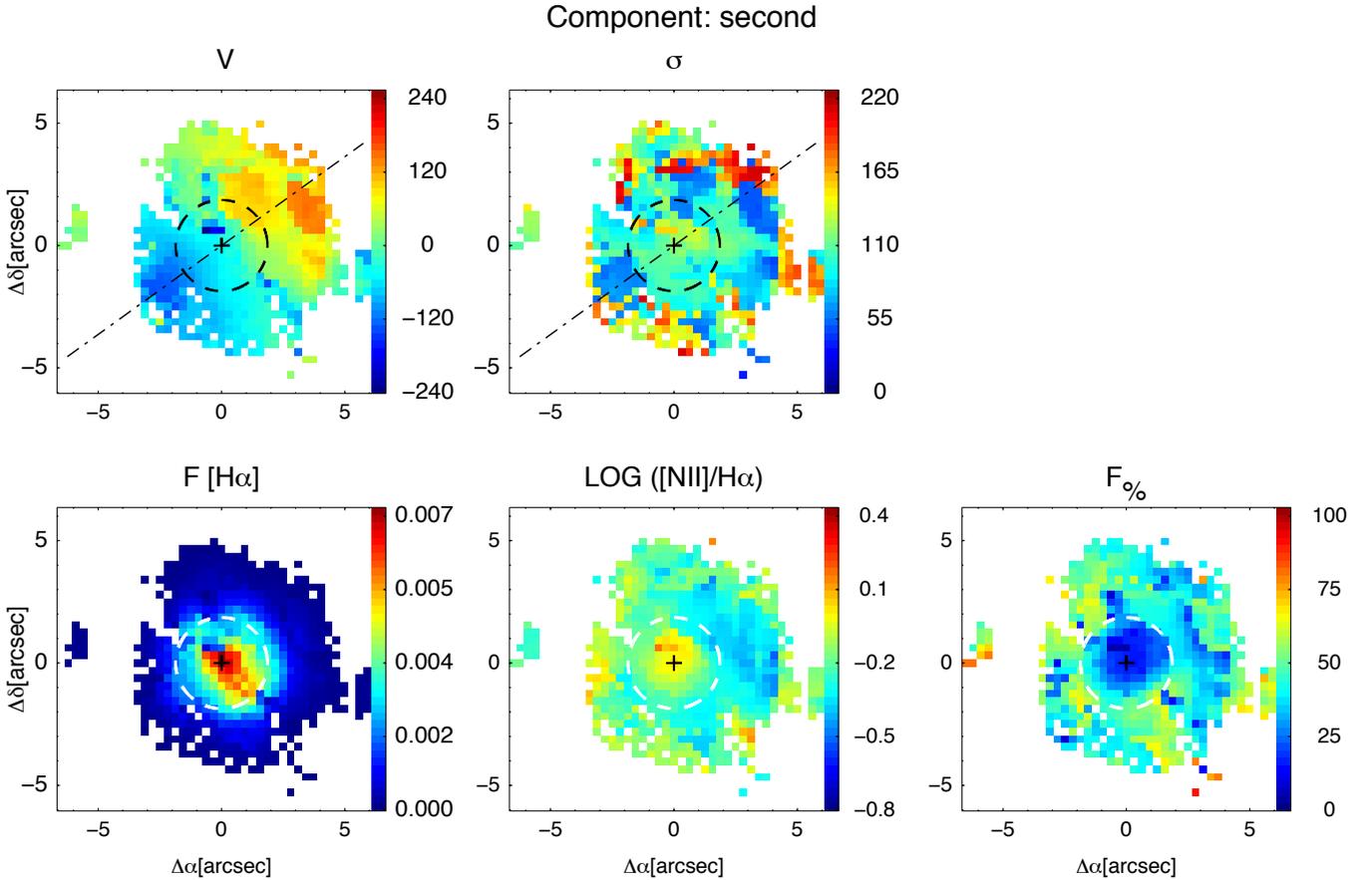}\\
\caption{The same as Fig.\,\ref{maps_1c}, but for the second component.}   
\label{maps_2c} 	
\end{figure*}

\begin{figure*}
\includegraphics[trim = 1.1cm .75cm 1.7cm 14.5cm, clip=true, width=1.\textwidth]{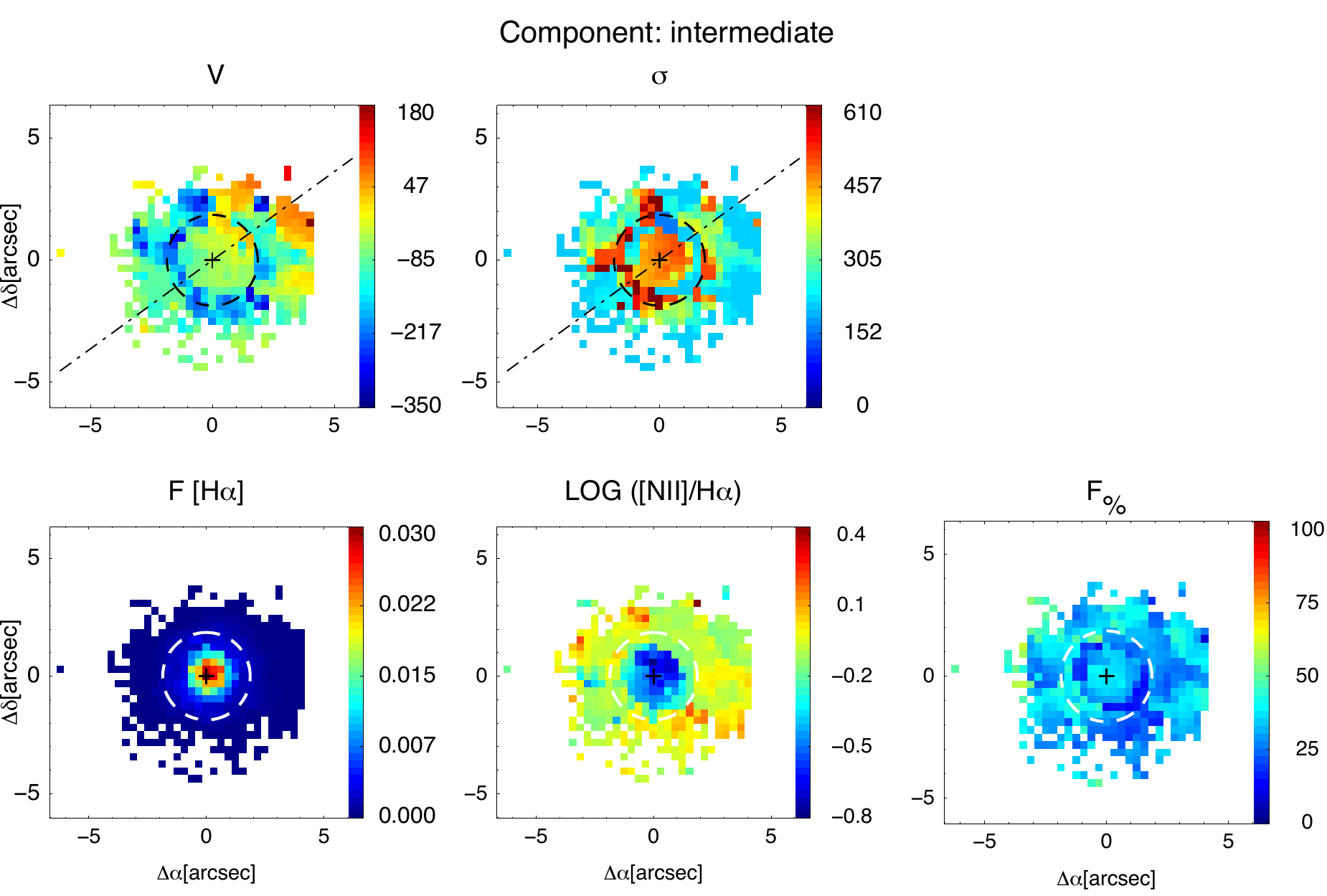}\\
\caption{The same as Fig.\,\ref{maps_1c}, but for the intermediate component.}    
\label{maps_3c} 	
\end{figure*}

\noindent Whereas the narrow and second components have a similar range of central wavelengths\footnote{Differences in the velocities of the two narrow components are of about 0.9\,\AA , i.e. $\sim$\,40\,km\,s$^{-1}$, on average. This value should not be considered as a systemic offset since the difference in velocity varies  throughout the MEGARA IFU field (Sections \ref{narrow} and \ref{second}; Figures \ref{maps_1c}, \ref{maps_2c} and \ref{pv_sigma}).}, a clear distinction  in line width exists, reaching up to a factor of 10 (on average was $\sim$3).  Hence, by first generating the map of the ratio between the widths of the two components, we were able to  differentiate the two narrow components. i.e. systemic (or narrow) and second (broader) components, according to their widths, and then map their properties (fluxes and velocities and the corresponding errors). Note that,  the ranges of values for the velocity dispersion of these two narrow components partially overlap. However, the overlap is not uniform within the the entire IFU field varying spatially. Therefore we cannot fix line widths of the narrow components and a spaxel-by-spaxel basis is appropriate. For intermediate and broad component a clear difference in width already exists after the line fitting is applied procedure as  mentioned in Sect.\,\ref{line_fitting}. \\
By  means of in house \textsc{idl} procedures, we generate maps of:  flux intensity, velocity field (V), and velocity dispersion ($\sigma$), for all  the four components. The kinematics-maps (velocity and velocity dispersion) are displayed in Figures \ref{maps_1c}, \ref{maps_2c} and \ref{maps_3c} along with those for the H$\alpha$ flux intensity (F\,[H$\alpha$]) and the percentage contribution  (F$_{\%}$) of each component to the total H$\alpha$-[N\,{\small II}] emission.
For each component used to model [N\,{\small II}] lines and narrow H$\alpha$, the maps of the ratio between [N\,{\small II}]$\lambda$6584/H$\alpha$ were also generated. The histograms of the distribution of the line ratios are presented in Fig.\,\ref{bpts_isto}. \\
The radial velocity maps are scaled to the systemic velocity defined as  \textit{c}\,$\times$\,\textit{z}\,, i.e. 4891.7\,km\,s$^{-1}$ (\textit{z} is the redshift,  Table\,\ref{T_properties}). The observed systemic radial velocity is 4839.9 (4800.2) km\,s$^{-1}$ being measured as  the velocity of the narrow  (second) component  at the brightest spaxel in the continuum image. This spaxel is marked with a cross in all the maps from the MEGARA datacube which  we assume to be the  nucleus of the galaxy (photometric center).  The  velocities measured at the position of the photometric center are similar (within less than 2 per cent)  to the assumed systemic velocity (\textit{c}\,$\times$\,\textit{z}). \\
Velocities are barycentric, we did not apply any correction for the heliocentric velocity which is of 21.4\,km\,s$^{-1}$ as determined with the \textsc{iraf} command \lq rvcorrect\rq . \\
All the maps are displayed using the plotting package \textsc{jmaplot}  by \cite{Maiz2004}.


\section[]{Kinematic Analysis}
\label{Kinematic_Analysis}

\subsection[]{Position Velocity Diagrams}
\label{PV_diagrams}
As we will discuss in Sections \ref{main_observational_results} and \ref{discussion}, the kinematics of both narrow and second component show signatures of a rotating disc. Therefore, from the kinematic maps of both components (Figures \ref{maps_1c} and \ref{maps_2c}, top panels) we extracted the velocity values in a 0.6\,arcsec pseudo slit along their major kinematic axis. To obtain the corresponding position-velocity diagrams (\lq PV-diagrams\rq), shown in top panels of Fig.\,\ref{pv_sigma}, we rescaled the velocities to the kinematic center. It is defined, for each component, as the velocity for which the rotation curve, Fig.\,\ref{pv_sigma} bottom panels, is symmetric.  For the narrow (second) component the velocity of the kinematic center is 4895.7\,km\,s$^{-1}$ (4874.2\,km\,s$^{-1}$) and it is located within 0.15\,arcsec (0.8\,arcsec) from the photometric center. The PV-curves for narrow and second components show distinct shapes (Sections \ref{narrow} and  \ref{second}) which will be discussed in detail in Sect.\,\ref{comparison_narrow}. At the same locations of the PV-curves, we extracted the corresponding velocity dispersion measurements. These will be discussed in comparison with the results of our disc modelling in Sect.\,\ref{disc_kinemetry}.

\subsection{Disc modelling with \textsc{kinemetry}}
\label{disc_kinemetry}

To obtain a  physically meaningful fit to the observed velocity field and velocity dispersion maps, and to robustly identify and quantify the possible deviations from an  ideal rotating-disc, we applied the  \textsc{kinemetry} package  (version 4.2) developed by \citet{Krajnovic2006} and methodology. Briefly, this procedure is an extension of surface photometry to the higher order moments of the velocity distribution. More specifically,  the \textsc{kinemetry} algorithm describes the data by a series of concentric ellipses of increasing major axis length. Along each ellipse, the moment as a function of angle is then extracted and decomposed into the Fourier series following the \textit{cosine}-law approximation (i.e. the velocity field in an ideal rotating disc is expected to be dominated by the \textit{cos}-term, \citealt{Shapiro2008}).\\
\noindent The disc-kinematic modelling  and the extraction of the kinemetric parameters is done with  \textsc{kinemetry} as follows. The first step is to locate the center of the system, around which the  ellipses are constructed. We assumed that the kinematic center is coincident with the photometric center (i.e. the peak of the continuum emission, Sect.\,\ref{maps_generation}). Other  relevant geometric parameters (i.e. position angle and inclination) will be derived during the \textsc{kinemetry} analysis at each ellipses's position (in steps of 0.31\,arcsec). Then, as the high SNR emission has an irregular morphology and it is not covering all the FoV (Sect.\,\ref{emlines_maps}), we relax the condition in the keyword \lq \textsc{cover}\rq \ which controls the radius at which the extraction of values from maps stops. We considered the value of 0.60 (default is 0.75) meaning that if less than the 60 per cent of the points along an ellipse are not present the procedure is interrupted. With these prescriptions, we generated the model map for the velocity field of the narrow component. This is shown in Fig. \ref{kinemetry_narrow_panels} (upper panels) where we show the observed and reconstructed maps as well as the residuals. In Fig.\,\ref{kinemetry_narrow_moments} the selected kinemetric coefficients relevant for the characterization of the disc kinematics are displayed as a function of the distance from the center of the system. These coefficients  are \lq $\Gamma$\rq : the kinematic PA;  and  \lq \textit{q}\rq : the flattening of the fitting-ellipse. Then for the velocity:  \lq \textit{k}$_{1}$\rq measures the amplitude of bulk motions (rotation curve) and  \lq \textit{k}$_{5}$\rq \ the deviations from a rotation disc pattern.  The kinematic PA has been also obtained independently ($\Gamma$$_{\rm\,FIT}$) using the method described in Appendix\,C of \citet{Krajnovic2006} via the \textsc{idl} routine \textsc{fit$\_$kinematic$\_$pa} (see also \citealt{Cappellari2007}).\\
As the velocity dispersion map of the narrow component deviates from what expected in the case of an ideal rotating disc (see Sect.\,\ref{narrow}), we consider to use \textsc{kinemetry} with an additional constraint. Specifically, the velocity dispersion map is reconstructed along the best-fitting ellipses from the modelling of the velocity field of the narrow component. The model map, the reconstructed map and the residuals are shown in Fig. \ref{kinemetry_narrow_panels} (lower panels). For velocity dispersion, the kinemetric coefficients considered are:  \lq \textit{A}$_{0}$\rq , the velocity dispersion profile with radius and  \lq \textit{B}$_{4}$\rq , the shape parameter (see Fig.\,\ref{kinemetry_narrow_moments}). The latter is used to quantify anisotropies, being the analogous of the photometric term that describes the deviation of the isophotes from an ellipse \citep{Krajnovic2006}. Specifically,  negative values are indicative of boxiness (triaxal structure,  slow rotation), while positive values of disciness (oblate,  fast rotators), see e.g. \citet{Emsellem2007} for further details. \\

\noindent  The velocity dispersion map of the narrow component (Fig.\,\ref{maps_1c} top right) shows an enhancement along the minor-axis  (green colors, $\sigma$\,$\geq$\,60\,km\,s$^{-1}$) as well as a decrement in a few peculiar regions of the IFU field (blue colors). These substructures possibly perturbe the disc kinematics, as we comment in Sect.\,\ref{narrow}  and discuss in Sect.\,\ref{disc_perturbations}.\\
In order to test if the results from \textsc{kinemetry} are influenced by these features, we also executed \textsc{kinemetry} for the velocity dispersion map of the narrow component masking  them out. Specifically, we first build a mask flagging individual spaxels in the velocity dispersion map of the narrow component  at the location of all substructures. Then we combined it, within  \textsc{kinemetry}, with the same set of parameters as in the previous disc-modelling. The results of this test  will be discussed in Sections \ref{narrow} and \ref{comparison_narrow} (see also Fig.\,\ref{kinemetry_narrow_moments}, orange circles). \\ 

\noindent The estimation of the dynamical ratio and scale height (V/$\sigma$ and \textit{h}$_{\rm\,z}$, respectively), defined and discussed in Sect.\,\ref{disc_support_hz}, for the second narrow component suggest the presence of a dynamically hotter and thicker disc. Therefore, we did not attempt to model with \textsc{kinemetry} (suited for thin discs) the velocity and velocity dispersion maps of the second component.\\
In Fig.\,\ref{rVrot} (upper panel), the rotation curve from \textsc{kinemetry} is compared to that from the symmetrization of the PV-curve (Fig.\,\ref{pv_sigma} bottom left) and those from previous works. Similarly, the velocity dispersion radial profile obtained from  \textsc{kinemetry}  (Fig.\,\ref{kinemetry_narrow_moments} green circles) and from the MEGARA map (obtained as the  PV-measurements, Sect.\,\ref{PV_diagrams}) are displayed (middle panel). We also show the dynamical ratio calculated along each ellipse (bottom panel;  see discussion in Sect.\,\ref{disc_support_hz}).
\section[]{Main observational results}
\label{main_observational_results}

In Table\,\ref{T_kin}, we summarize the kinematic parameters and characteristic values of the [N\,{\small II}]/H$\alpha$ ratio and the contribution of the four components considered in the present paper (Sect.\,\ref{line_fitting}). \\

\subsection{Narrow (systemic) component}
\label{narrow}

The maps of the ionised gas kinematics as traced by the H$\alpha$-[N\,{\small II}] lines (narrow component) are shown in Fig.\,\ref{maps_1c} (top panels). The velocity field (Fig.\,\ref{maps_1c}, top left) presents a point-antisymmetric velocity pattern consistent with large kiloparsec-scale ordered rotational motions (e.g. a rotating disc).  The peak-to-peak semi-amplitude of the velocity field  is 163\,$\pm$\,1\,km\,s$^{-1}$ ($\Delta$V, Table\,\ref{T_kin}), with the maximum velocity gradient oriented  in the  south\,east  - north\,west direction with a position angle PA$_{\rm\,maps}$\,=\,(125\,$\pm$\,10)$^{\circ}$ (Table\,\ref{T_kin}). This value is in remarkably good agreement with the PA of the photometric major axis (PA$_{\rm\,phot}$) from Hyperleda, i.e. 126$^{\circ}$ (Table\,\ref{T_properties}).\\
The velocity dispersion map (Fig.\,\ref{maps_1c}, top right) shows an irregular pattern and it is not centrally-peaked, contrary to what is expected for a rotating disc. The maximum value  of the velocity dispersion map (108\,$\pm$\,4\,km\,s$^{-1}$) is in fair positional agreement (within 0.6\,arcsec, i.e. 200\,pc) with the nucleus. The average velocity dispersion inside the nuclear region (i.e. $\sigma_{\rm\,c}$,  40\,$\pm$\,1\,km\,s$^{-1}$, Table\,\ref{T_kin}) is rather similar to that  in the  putative disc  (i.e. $\sigma_{\rm\,avg}$, 38\,$\pm$\,1\,km\,s$^{-1}$, Table\,\ref{T_kin}). In this comparison, it has to be noted that there are some irregularities. Specifically, there are eight regions (six in the main disc and two at \textit{r}\,$>$\,5.0\,arcsec) of low velocity dispersion (i.e. $\sigma$\,$<$\,30\,km\,s$^{-1}$).  Furthermore, the velocity dispersion is enhanced (typically $\sigma$\,$>$\,60\,km\,s$^{-1}$) along the minor axis of the putative disc rotation towards north-east and south-west direction, up to \textit{r}\,$\sim$\,3.7\,arcsec. All the  anomalies of the velocity dispersion maps will be discussed  in Sect.\,\ref{disc_perturbations}.  These features are identified in the maps shown in Fig.\,\ref{regions}.

\noindent  For the narrow component, the regular pattern of the velocity map (Fig.\,\ref{maps_1c}, top left) indicates a relatively undisturbed velocity field without any strong signature of non-circular motions. The corresponding PV-curve is nearly symmetric (Fig.\,\ref{pv_sigma}  top left). It seems to level out at $\sim$\,150\,km\,s$^{-1}$, after a well-resolved and steep rise in the receding side (positive velocities) in the north-west direction. The approaching side (negative velocities, to the south-east) is slightly more irregular showing a discontinuos rise.  \\

\noindent The models produced with \textsc{kinemetry} (Sect.\,\ref{disc_kinemetry}) satisfactorily reproduce the data, in particular, for the velocity (Fig.\,\ref{kinemetry_narrow_panels} top).  Indeed, for the majority of the spaxels  (i.e. 63 per cent) in the residual map, data\,$-$\,model  (Fig.\,\ref{kinemetry_narrow_panels} top right), the values are lower than 15\,km\,s$^{-1}$. Larger negative residuals ($>$\,50\,km\,s$^{-1}$) are seen in correspondence of the minor axis of the rotation. \\
The reconstructed  velocity dispersion map is not centrally peaked showing one \lq drop\rq \ of $\sim$\,25\,km\,s$^{-1}$ (Fig.\,\ref{kinemetry_narrow_panels} bottom). This  feature is observed  at  \textit{r}\,$\leq$\,1.5\,arcsec. It is distinctly visible in the velocity dispersion radial profile in Fig.\,\ref{kinemetry_narrow_moments} (right) either when including (green circles) or excluding (orange circles) possible  velocity dispersion substructures (see Sect.\,\ref{disc_kinemetry}). Such a velocity dispersion decrement is possibly related to a $\sigma$-drop (the velocity dispersion is depressed in the galaxy center) as discussed in Sect.\,\ref{comparison_narrow}.\\

\noindent  When comparing the results from the \textsc{kinemetry} modelling including/excluding possible velocity dispersion anomalies some variations are seen for the  \textit{B}$_{4}$ parameter. In the majority of the cases (8 out of 13 measurements) the shape of the ellipses is preserved (either disky or boxy) being the \textit{B}$_{4}$\,/\,\textit{A}$_{0}$ ratio  the same within uncertainties in half of these cases (4 out of 8). For the other five cases, the  \textit{B}$_{4}$\,/\,\textit{A}$_{0}$ ratios have different signs indicating either the disciness or boxiness of the ellipses (Fig.\,\ref{kinemetry_narrow_moments} bottom-right). The instability of the \textit{B}$_{4}$ term could be due to the limitation of the parameters space for \textsc{kinemetry}, e.g. for a given input ellipse, since the number of points is lower due to the masking. From this comparison, we excluded the values at \textit{r}\,$>$\,4.8\,arcsec as, when velocity dispersion anomalies are masked out,  at this radius \textsc{kinemetry} halts (see Sect.\,\ref{disc_kinemetry}).\\

\noindent The position angle from \textsc{kinemetry}, $\Gamma$, is remarkably stable with values within 120$^{\circ}$ and 130$^{\circ}$  over the most of the sampled radius (with only 3 exceptions, Fig.\,\ref{kinemetry_narrow_moments} top left).  The average value of $\Gamma$ ($\sim\,$125$^{\circ}$) is consistent with those obtained with different estimates, i.e. from a direct estimate using the observed maps PA$_{\rm\,maps}$\,=\,(125\,$\pm$\,10$^{\circ}$, Table\,\ref{T_kin}) and from the output of the  \textsc{fit$\_$kinematic$\_$pa} routine (Sect.\,\ref{disc_kinemetry}), i.e. $\Gamma$$_{\rm\,FIT}$\,=\,(124.3\,$\pm$\,2.5$)^{\circ}$.\\
The \textit{q} parameter has values between  0.4 and 1 (Fig.\,\ref{kinemetry_narrow_moments} top right).  The average (median) value of  \textit{q} is 0.88 (0.94), indicating that the modeled velocity map for the  narrow component is kinematically round. In the special case of a disc where the motions are confined to the plane (thin-disc approximation, see also Sect.\,\ref{disc_support_hz}) the flattening is directly related to the inclination of the disc, i.e. \textit{q}\,=\,cos(\textit{i}) \citep{Krajnovic2006}. This would imply an inclination of 31.5$^{\circ}$, that is consistent within 1 per cent with the value reported in Hyperleda (i.e. 30.2$^{\circ}$, Tab.\,\ref{T_properties}).\\

\begin{figure*}
\includegraphics[trim = 2.15cm 16.95cm 2cm 6.7cm, clip=true, width=1.\textwidth]{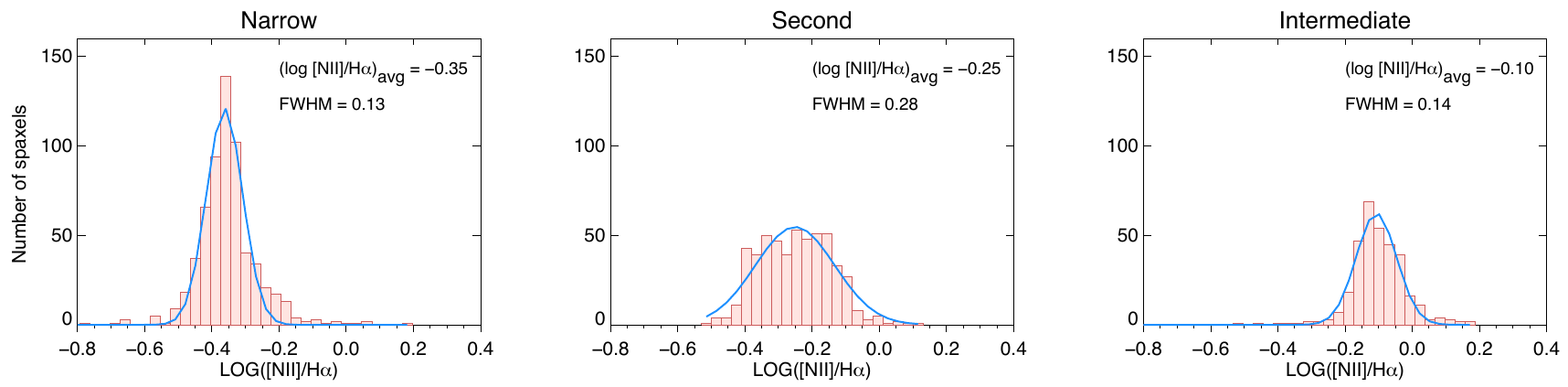}\\
\caption{Distribution of the logarithm of the [N\,{\small II}]/H$\alpha$ line ratio for the three components (indicated at the top of each panel) used to model [N\,{\small II}] lines and narrow H$\alpha$ during this work (Sect.\,\ref{line_fitting}). We only considered the measurements outside the nuclear region (dashed circle in all spectral maps). As reference, in all  panels, the Gaussian fit to the distribution of the values for the wavelength and width is marked with a continuous blue line. Average values and the FWHM of the Gaussian fit are  reported on the top right. All histograms are binned in 0.03 dex-wide.}
\label{bpts_isto} 	
\end{figure*}

\noindent The map of the H$\alpha$ flux-intensity (F[H$\alpha$]) and that of the percentage contribution to the total H$\alpha$-[N\,{\small II}] emission (F$_{\%}$) in Fig.\,\ref{maps_1c} (bottom panels) show different characteristics. On one hand, the H$\alpha$ flux-intensity shows  two regions with enhanced flux  at nearly symmetric positions with respect to the minor photometric axis (north-east to south-west direction). These roughly correspond to the brightest star-forming clumps seen in the \textit{HST} image  (Fig.\,\ref{HST_morphology} middle).  On the other hand, the map of F$_{\%}$ is depressed in the nuclear region (F$_{\%, c}$\,=\,12\,$\pm$\,9, Table\,\ref{T_kin}) with respect to that of the main disc (F$_{\rm\,\%,\,avg}$\,=\,46\,$\pm$\,24, Table\,\ref{T_kin}).\\
The map of the  log\,([N\,{\small II}]/H$\alpha$) is rather homogenous (Fig.\,\ref{maps_1c} bottom) except two evident substructures. One is located towards the south-west with negative values, i.e. log\,([N\,{\small II}]/H$\alpha$)\,$\sim$\,$-$\,0.6, roughly at the same location of one of the regions of low velocity dispersion (region 6  in Fig.\,\ref{regions}). The other is along the minor axis of the putative disc rotation towards north-east and south-west direction (partially overlapping the regions where velocity dispersion is enhanced, Fig.\,\ref{maps_1c} top-right) and there log\,([N\,{\small II}]/H$\alpha$) is $\sim$\,0.1. Outside the nuclear region, the distribution of the line ratios has an average value $-$\,0.35 (Fig.\,\ref{bpts_isto} left); the standard deviation is 0.14 (Table\,\ref{T_kin}).

\subsection{Second narrow component}
\label{second}

\noindent The kinematics (velocity and velocity dispersion) of the second component (Fig.\,\ref{maps_2c}, top panels) have a complex shape, but, overall, these maps suggest the presence of an irregular disc. Specifically,  the ionised gas velocity map of the second narrow component seems to show both blue and red sides similarly to the spider-pattern expected in the case of a rotation-dominated system. The deviations from an ideal rotating disc include poorly defined kinematic axes, and asymmetric velocity distribution in the receding side (positive velocities) with respect to the, more regular, approaching side (negative velocities) of the putative disc. 
The differences in the \lq regularity\rq \  between the two sides of the disc are  visible in the symmetrised rotation curve in Fig.\,\ref{pv_sigma} (bottom right). For example, the drop of about 50\,km\,s$^{-1}$ is visible only in the receding side. Moreover, contrary to what it is expected for a rotating disc, the velocity dispersion is not centrally peaked, with evident  irregularities at large radii, i.e. \textit{r}\,$>$\,3.5\,arcsec to the north-west. The distribution of the velocity dispersions is flat with $\sigma_{\rm\,avg}$\,$\sim$\,$\sigma_{\rm\,c}$ (Table\,\ref{T_kin}) but with the lowest values ($<$\,40\,km\,s$^{-1}$) at a \textit{r}\,$>$\,2\,arcsec. Interestingly, five out of the eight regions  with the lowest velocity dispersion values for the narrow component  (Sect.\,\ref{narrow}) seem to have a counterpart also in the velocity dispersion map of the second component where  $\sigma$\,$<$\,50\,km\,s$^{-1}$ (Fig.\,\ref{maps_2c}, top right).\\

\noindent As mentioned above, the velocity field shows some perturbations (particularly in the red-side with positive velocities, Figures \ref{maps_2c} top left and \ref{pv_sigma} bottom right) deviating from the putative rotating disc-like pattern. Nonetheless, the PV-curve is  nearly symmetric flattening at 140\,km\,s$^{-1}$ (Fig.\,\ref{pv_sigma}, top right), after a shallow rise. Note that this curve has been calculated after excluding the three spaxels with  very blue shifted velocities in the nuclear region, as these clearly deviate from the main rotation-like pattern of the velocity field (due to a not fully satifactory line modelling, Sect.\,\ref{line_fitting}).\\

\noindent The H$\alpha$ flux-intensity is rather homogenous and centrally peaked.  The dominating feature in the  percentage contribution map  is the sharp drop (of about 15 per cent) toward the nuclear region (Table\,\ref{T_kin}). \\
The map of  log\,([N\,{\small II}]/H$\alpha$) (middle lower panel in Fig.\,\ref{maps_2c}) is less homogenous than the corresponding H$\alpha$ flux map (Fig.\,\ref{maps_2c}, bottom left). The lowest values, i.e. log\,([N\,{\small II}]/H$\alpha$)\,$\sim$\,$-$\,0.4, are concentrated in the north-west area. Outside the nuclear region, the distribution of the line ratios has an average  value $-$\,0.25 (Fig.\,\ref{bpts_isto} center); the standard deviation is 0.11 (Table\,\ref{T_kin}).

\begin{table*}
\caption{Main properties of the different components of NGC\,7469.} 
\begin{tabular}{l c c c c c c c c}
\hline
Component  & $\Delta$V & PA$_{\rm\,maps}$  & $\sigma_{\rm\,c}$ & $\sigma_{\rm\,avg}$  & F$_{\rm\,\%,\,c}$ & F$_{\rm\,\%,\,avg}$  & log\,([N\,{\small II}]/H$\alpha$)$_{\rm\,c}$ & log\,([N\,{\small II}]/H$\alpha$)$_{\rm\,avg}$ \\
 &km\,s$^{-1}$ &degree& km\,s$^{-1}$ & km\,s$^{-1}$ & & \\  
\hline
Narrow         &  163\,$\pm$\,1 & 125\,$\pm$\,10 & 40\,$\pm$\,1 (15) & 38\,$\pm$\,1 (12) & 12\,$\pm$\,9 & 46\,$\pm$\,24 & $-$\,0.32\,$\pm$\,0.01 (0.11) & $-$\,0.35\,$\pm$\,0.02 (0.14)\\
Second         &  137\,$\pm$\,2 & 120\,$\pm$\,15 & 101\,$\pm$\,2 (21) &108\,$\pm$\,4 (42) & 29\,$\pm$\,12 & 44\,$\pm$\,12 & $-$\,0.18\,$\pm$\,0.01 (0.11) & $-$\,0.25\,$\pm$\,0.02 (0.11) \\
Intermediate &   ---  &  --- & 388\,$\pm$\,5 (137) & 276\,$\pm$\,8 (100) & 29\,$\pm$\,8 & 32\,$\pm$\,8 &   $-$\,0.34\,$\pm$\,0.02 (0.21)  &   $-$\,0.10\,$\pm$\,0.02 (0.10)  \\
Broad 	    &   ---  &  ---  & 1100\,$\pm$\,10 (156) & --- & 41\,$\pm$\,7 & ---  & ---  & --- \\
\hline
\end{tabular}
\label{T_kin}
\begin{flushleft}
{\textit{Notes.}  Summary of the main kinematics properties. \lq $\Delta$V\rq \ is the peak to peak semi-amplitude of the velocity field.  \lq PA$_{\rm\,maps}$ \rq \ is the position angle of the kine- matic  major axis, as measured from the north (anti clockwise), in the MEGARA maps. \lq$\sigma_{\rm\,c}$\rq \ (\lq$\sigma_{\rm\,avg}$\rq) is the average velocity dispersion calculated from the observed velocity dispersion in (excluding) the nuclear region (Sect.\,\ref{observations_datared}). Similarly, but considering the percentage-fraction (F$_{\%}$) and logarithm of [N\,{\small II}]/H$\alpha$, in the last columns. In parenthesis, we report the standard deviation values. Uncertainties correspond to the median value in the error-maps except for F$_{\%}$ for which it coincides with the  standard deviation.}
\end{flushleft}
\end{table*}

\subsection{Intermediate component}
\label{intermediate}

The velocity  and velocity dispersion maps of the intermediate component, shown in the top panels of Fig.\,\ref{maps_3c},  are irregular, somewhat chaotic,  and lack of any rotating-disc feature. \\

\noindent  Overall, the velocity distribution is skewed to negative (blueshift) velocities. Combining the informations from both kinematic maps (velocity and velocity dispersion), we can characterized three spatially distinct regions  with different kinematic properties. Outside the nuclear region, on the one hand, a region at nearly the systemic velocity, $\pm$\,100\,km\,s$^{-1}$, with a moderately high velocity dispersion ($<$\,300\,km\,s$^{-1}$). On the other hand,  extreme blue shifted velocities (V$<$\,-200\,km\,s$^{-1}$) correspond to an enhancement of velocity dispersion ($>$\,400\,km\,s$^{-1}$). The former has an irregular morphology and it is observed at a distance of $\sim$\,2.5\,arcsec from the photometric center, while the latter has a ring-like morphology with a radius of $\sim$\,1.7\,arcsec.  Leaving aside these two regions, the nuclear region (see Sect.\,\ref{observations_datared}) shows an intermediate kinematics with respect to the other two regions, with rest frame velocities (between $-$\,90 and 10\,km\,s$^{-1}$)  but high velocity dispersion, 388\,$\pm$\,5\,km\,s$^{-1}$ (Table\,\ref{T_kin}). \\

\noindent The H$\alpha$ flux-intensity is  strongly centrally peaked (Fig.\,\ref{maps_3c} bottom left). The map of F$_{\%}$ (Fig.\,\ref{maps_3c} bottom right) is rather regular (nearly $\sim$\,30 per cent in the entire FoV, Table\,\ref{T_kin}) in comparison to that of narrower components (Figures \ref{maps_1c} and \ref{maps_2c}). \\
The map of the  log\,([N\,{\small II}]/H$\alpha$) is not homogenous but not clear structures are seen (Fig.\,\ref{maps_3c} bottom). Outside the nuclear region the distribution of the line ratios has an average value $-$\,0.10 (Fig.\,\ref{bpts_isto} right); the standard deviation is 0.10 (Table\,\ref{T_kin}).   


\subsection{Broad component}
\label{Broad}
The broad H$\alpha$ component is the dominating component in the nuclear region where F$_{\%,\,c}$\,=\,41\,$\pm$\,7 (Table\,\ref{T_kin}). It is  originated in the unresolved AGN's BLR. \\
For this component, we measured velocities that are typically rest frame, within $\pm$\,90\,km\,s$^{-1}$, with respect to the systemic. Overall, the average velocity dispersion is $\sigma_{\rm\,c}$\,=\,1100\,$\pm$\,10\,km\,s$^{-1}$ (Table\,\ref{T_kin}), i.e. $\sim$\,2590\,km\,s$^{-1}$ in FWHM.

\begin{figure*}
\includegraphics[trim = .15cm 11.5cm 2.05cm 2.65cm, clip=true, width=1.\textwidth]{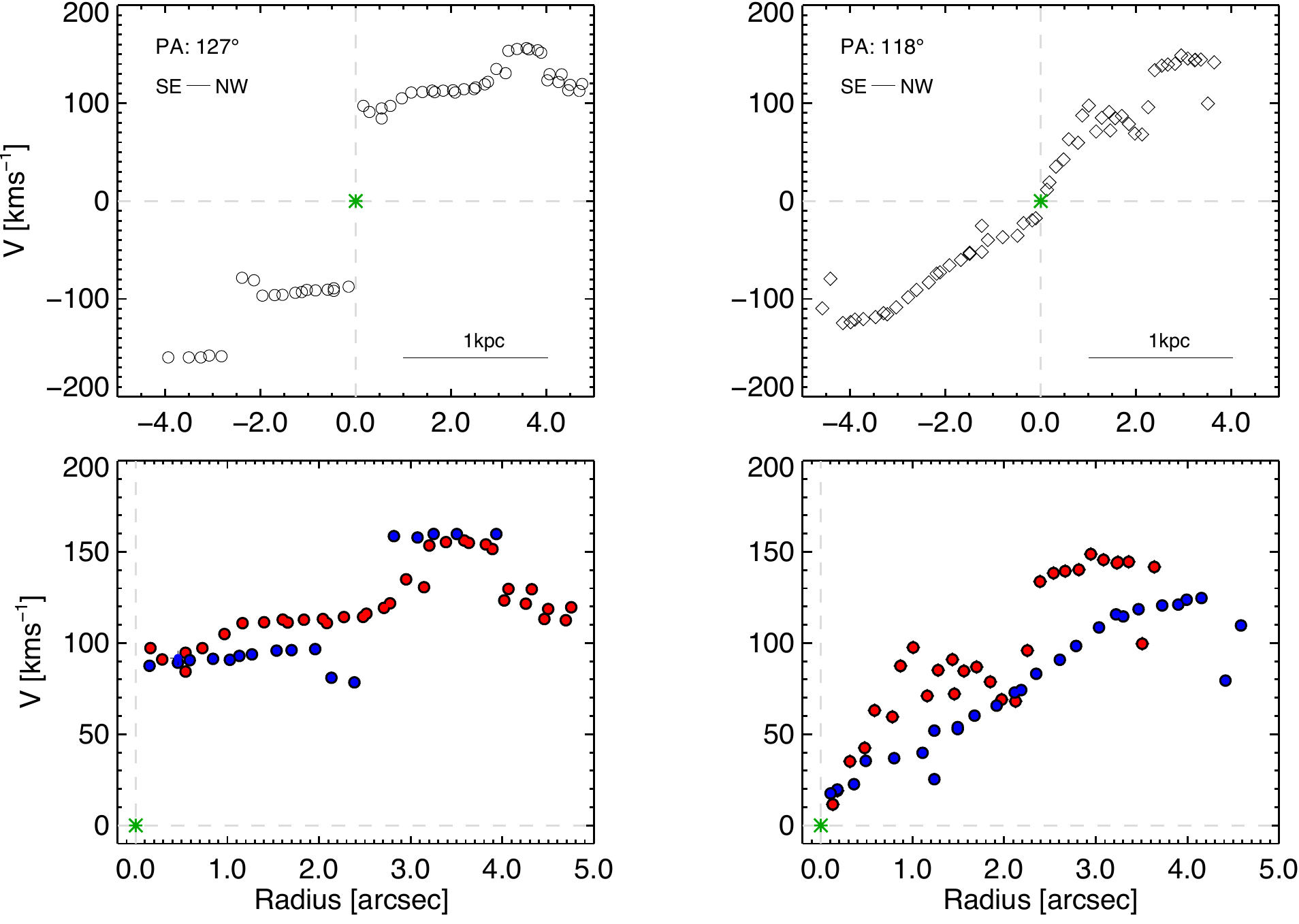}\\
\caption{PV- and rotation curves for the narrow (left) and the second (right) kinematic components, upper and lower panels respectively. PV-curves were obtained considering a pseudo-slit aligned according to the major axis of their rotation. Circles and diamonds indicate the point for the narrow  and second  kinematic components, respectively. 
The radius is calculated as the distance from the kinematic center that is marked with a green asterisk (see Sect.\,\ref{PV_diagrams}).  In the bottom panels, blue and red symbols indicate the approaching (negative velocities) and receding sides (positive velocities) of the rotation, respectively. In all panels, gray dashed lines show zero-points for position and velocity, as reference. The typical uncertainty on the velocity measurements is generally  $\leq$\,8\,km\,s$^{-1}$. Hence, error bars are not showed as being comparable with the symbols size.}   
\label{pv_sigma} 	
\end{figure*} 
\section[]{Discussion}
\label{discussion}

\begin{figure*}
\includegraphics[trim = 1.0cm .75cm 1.5cm 15.4cm, clip=true, width=.94\textwidth]{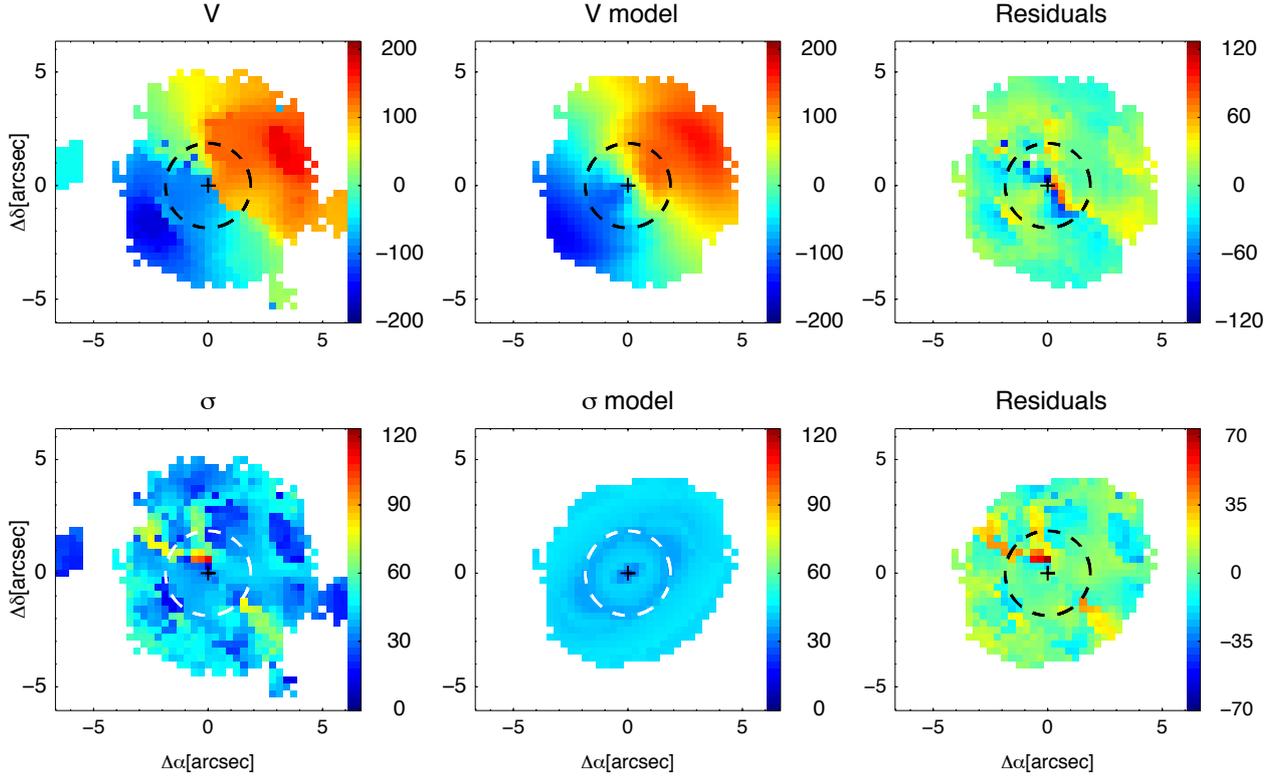}\\
\caption{For the narrow component, the maps of the H$\alpha$ velocities (top) and velocity dispersion (bottom) and their respective reconstructed (middle) and residual (data\,$-$\,model, right) maps  for NGC\,7469 (se also Sect.\,\ref{disc_kinemetry}). The cross marks the photometric center and  the dashed  circles indicates the nuclear region (Sect.\,\ref{observations_datared}). All the maps are in km\,s$^{-1}$ units.}
\label{kinemetry_narrow_panels} 	
\end{figure*} 
\begin{figure*}
\includegraphics[trim = 2.2cm 21.2cm 1.75cm 2.65cm, clip=true, width=.970\textwidth]{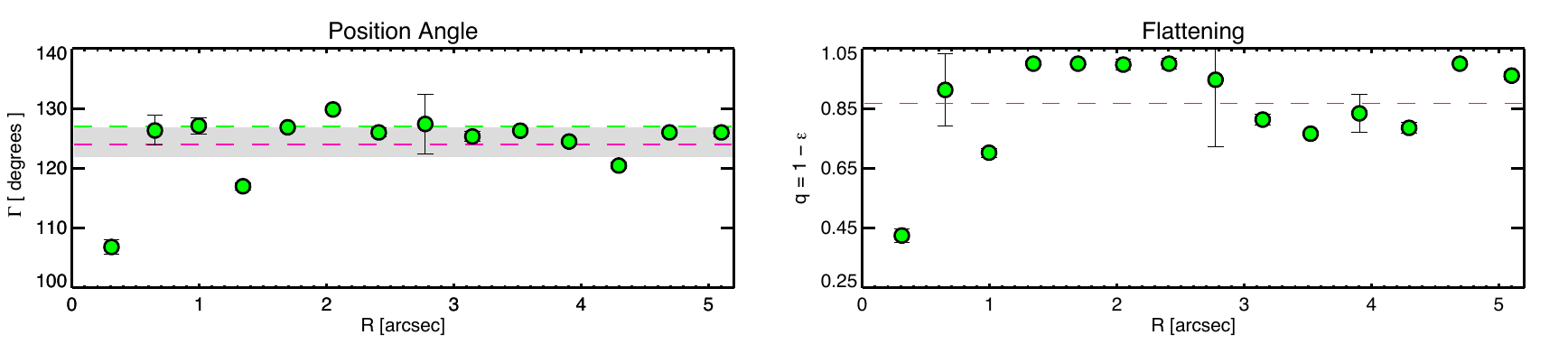}\\
\includegraphics[trim = 2.2cm 18.65cm 1.75cm 2.9cm, clip=true, width=.965\textwidth]{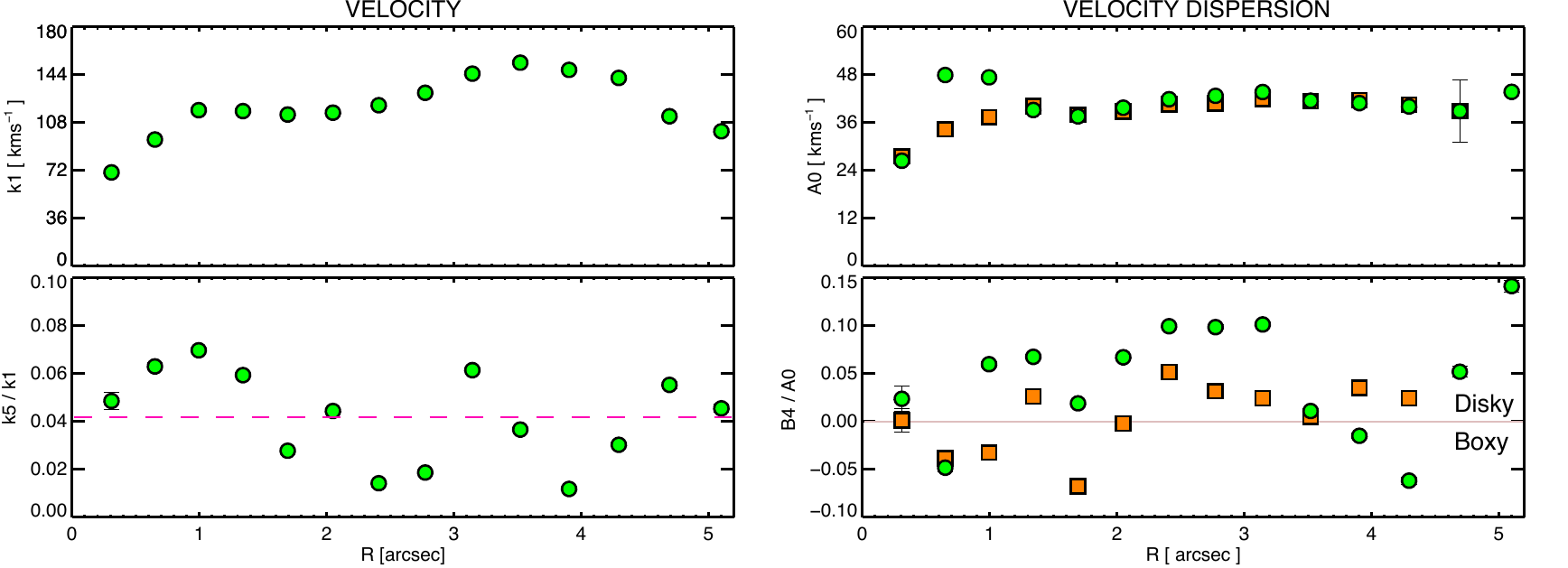}\\
\caption{Green circles mark the radial profiles of the kinemetric parameters describing the model in Fig.\,\ref{kinemetry_narrow_panels}. In the upper panels,  the position angle $\Gamma$ and the flattening  \textit{q} of the best-fitting ellipses ($\varepsilon$ is the ellipticity) are shown. In the lower panels, on the left, we display the first  (\textit{k}$_{1}$) and fifth order (\textit{k}$_{5}$) of the Fourier terms, as well as,  on the right, the intensity and shape parameter (\textit{A}$_{0}$ and \textit{B}$_{4}$, respectively). In the top left panel,  magenta and green dashed lines indicate the average values of  $\Gamma$ and  PA$_{\rm\,maps}$ (i.e. the major axis of the rotation from observed maps, Fig.\,\ref{maps_1c}, Table\,\ref{T_kin}) respectively.  In grey, the range of values for $\Gamma$$_{\rm\,FIT}$\,=\,(124.3\,$\pm$\,2.5)$^{\circ}$ resulting from the method for calculating the PA by  \citet{Krajnovic2006} (see Sect.\,\ref{disc_kinemetry}). The value for the photometric PA, PA$_{\rm\,phot}$ (Table\,\ref{T_properties}), is not marked but is in good agreement with the value of PA$_{\rm\,maps}$ (Table\,\ref{T_kin}). In the bottom left panel, the dashed magenta line marks the average value of  \textit{k}$_{5}$/\textit{k}$_{1}$ ($\sim$\,0.04, see \citealt{Krajnovic2011}). In the bottom-right panel, the light pink line indicates the zero, which is also the dividing line indicating  the deviations of the iso-$\sigma$ contours from an ellipse, i.e. boxiness and disciness (negative and positive values respectively, \citealt{Krajnovic2006}). Orange squares mark the radial profiles of the kinemetric parameters from the test  described in Sect.\,\ref{disc_kinemetry} for which we excluded all the anomalies in the velocity dispersion map (Sections \ref{narrow} and  \ref{disc_perturbations}). Note that, due to this masking, \textsc{kinemetry} halts at \textit{r}\,$\sim$\,4.8\,arcsec. In all panels, uncertainties are the formal (1\,$\sigma$) errors of the coefficients which are returned by \textsc{kinemetry}. }  
\label{kinemetry_narrow_moments} 	
\end{figure*} 

\subsection{Comparison with previous H$\alpha$ broad  component measurements}
\label{prop_BLR}
Most of the previous optical studies of the BLR in NGC\,7469 are focussed on measurements of the FWHM of the H$\beta$$\lambda$4861 line from reverberation mapping  campaigns (e.g. \citealt{Peterson2004, Peterson2014}) with only a few direct estimates (e.g. \citealt{Du2014}). These works provide a range of the FWHM(H$\beta$)   between 1967\,km\,s$^{-1}$ \citep{Du2014} and 2169\,$\pm$\,459\,km\,s$^{-1}$ \citep{Peterson2014}. The unique measurement of the FWHM of the, less studied, H$\alpha$ broad component  is 1615\,$\pm$\,119\,km\,s$^{-1}$ from long-slit spectroscopy data at spectral resolution of $\sim$\,9\,\AA \ (\textit{\lq AGN Watch project\rq}, \citealt{Peterson2004} and \citealt{Bentz2015} and references therein). All these previous measurements are narrower compared to that from MEGARA data in the present work, i.e. $\sim$\,2590\,km\,s$^{-1}$ on average (see Sect.\,\ref{Broad}). This could be explained in terms of spectral resolution. The unprecedented spectral resolution of MEGARA allows a multi-component modelling of the H$\alpha$-[N\,{\small II}] blend (not accounted for in previous works), that results in a more accurate estimation of the line parameters, such as the FWHM, of each component, including the broad one from the BLR. Another possibility for explaining the differences in the FWHM of the broad component is AGN variability. \citet{Pronik2009} reviews a compilation of published data along with  those available in literature  (1975\,-\,2006) for NGC\,7469 showing flux variability of  the BLR (in time scales of $\sim$20 days or less, see also \citealt{Baldi2015} and \citealt{Pronik1976}). On the basis of  the dramatic spectral-variability of the optical emission line profiles (fading and raising of  the H$\beta$ broad component), NGC\,7469 has been classified as \lq changing-look\rq \ AGN by  \citet{Chuvaev1990}.\\ 
\noindent In order to address the variability scenario, we degraded the  nuclear spectrum, i.e. that extracted at the photometric center, from the MEGARA datacube (at reciprocal linear dispersion of $\sim$\,0.0974\,\AA/pixel, Sect.\,\ref{observations_datared}) to the resolution of $\sim$\,9\,\AA \ \citep{Peterson2004,Bentz2015}. By modelling the H$\alpha$-[N\,{\small II}] emission  in the new low-resolution nuclear spectrum with two Gaussian components (narrow\,+\,broad, as in Sect.\,\ref{line_fitting}), we found the broad component to have a FWHM of about 2100\,km\,s$^{-1}$ which is consistent with the variability hypothesis.\\

\noindent Thanks to MEGARA observations, we also provide the first estimate for the percentage contribution of the  broad AGN component to the total H$\alpha$-[N\,{\small II}] profile (i.e. 41\,$\pm$\,7 per cent, Table\,\ref{T_kin}) which is dominant in the nuclear region. 


\subsection{Disc kinematics}
\label{comparison_narrow}

\noindent  For both narrow components, the kinematic  major axes are well aligned with that of the photometric axis as seen in the \textit{HST} images (Fig.\,\ref{HST_morphology}) with  PA$_{\rm\,maps}$\,$\sim$\,PA$_{\rm\,phot}$ (Tables \ref{T_properties} and \ref{T_kin}, Sect.\,\ref{narrow}).\\

\noindent  Although the velocity fields of the two narrow components show an overall similar rotation pattern, with blueshifts to the south-east and redshifts to the north-west  (Figures \ref{maps_1c}, \ref{maps_2c} and \ref{pv_sigma}), it can be clearly seen that the two components present somewhat distinct kinematics in terms of velocity amplitude  and velocity dispersion (Table\,\ref{T_kin}). Moreover, the PV- and rotation curves for the narrow and the second components show distinct shapes (Fig.\,\ref{pv_sigma}) suggesting possible lags in velocities, besides the peak-to-peak velocities are similar (Table\,\ref{T_kin}). In terms of the velocity dispersion, for both narrow components the spectral maps (Figures \ref{maps_1c} and \ref{maps_2c} top right) are  irregular and not centrally peaked, hence they deviate from what is expected in  the case of  rotating discs (Sect.\,\ref{narrow}). \\

\noindent  For the narrow component, the rotation curves  derived from our MEGARA data in Fig.\,\ref{rVrot} (top panel, blue triangles and red circles)  indicates a good agreement  with those from  past works  although  the differences in  the spatial coverage and ISM gas phases considered. Indeed, previous studies of NGC\,7469 at optical, radio and NIR wavelengths have shown an overall rotation-dominated kinematics of the ionised and molecular gas \citep{Fathi2015, MullerSanchez2011, Hicks2009}. Nevertheless, asymmetries and perturbations of the rotation curve of the galaxy and the existence of distinct kinematics components have been also observed (e.g. \citealt{Marquez1994, Davies2004}). In Sect.\,\ref{kin_prev}, a detailed comparison between the results from present MEGARA IFS data and previous works is presented. \\
  Overall, the  velocity field of the narrow component has intermediate properties between \textit{regular rotators} (i.e. \textit{k}$_{5}$/\textit{k}$_{1}$\,$<$\,0.04) and \textit{non regular rotators} (i.e. \textit{k}$_{5}$/\textit{k}$_{1}$\,$>$\,0.04) according to the scheme proposed by \citet{Krajnovic2011} with an average value of \textit{k}$_{5}$/\textit{k}$_{1}$  of $\sim$\,0.04. Moreover,  there is not indication of a kinematic twist (the standard deviation of the values of $\Gamma$ is $<$\,10$^{\circ}$) or other complex kinematic features such as counter rotating cores or kinematically distinct cores  \citep{Krajnovic2011}. \\

\noindent For the narrow component, the average velocity dispersion of the disc,  38\,$\pm$\,1\,km\,s$^{-1}$ (Table\,\ref{T_kin}), is in fair agreement with the spectroscopic measurements  by \citet{Epinat2010}, i.e. $\sigma$\,$\sim$\,15-30\,km\,s$^{-1}$ (GHASP survey) suggesting that NGC\,7469 share a similar dynamical status with local spirals. Although, this is not the case for the second component. The   average velocity dispersion of the disc, 108\,$\pm$\,2\,km\,s$^{-1}$ (Table\,\ref{T_kin}),    indicate a much thicker disc than in normal spirals or than the one traced by the narrow component being also larger than the typical value found at high-z, i.e. $\sigma$\,$\sim$\,60-90\,km\,s$^{-1}$ \citep{ForsterSchreiber2011}.\\

\noindent The reconstructed \textsc{kinemetry} map (Fig.\,\ref{kinemetry_narrow_panels} right)  is  rather flat  displaying a drop of $\sim$\,25\,km\,s$^{-1}$ in the innermost region (\textit{r}\,$\leq$\,1.5\,arcsec). The velocity dispersion profile from MEGARA data  shown in the middle panel of Fig.\,\ref{rVrot} (red circles)  is rather flat with values between 15 and 50 km\,s$^{-1}$. The two main features are: the absence of a central peak (expected for regular rotating discs) and the decrease at $\sim$\,3.3\,arcsec from the nucleus. The former feature is discussed in what follows  along with the results from our disc modelling (Sect.\,\ref{disc_kinemetry}). The latter feature is due to the presence of two regions with low velocity dispersion (see also Fig.\,\ref{maps_2c} top) located along the major axis of the rotation (hence within the psuedoslit, Sect.\,\ref{PV_diagrams}). These are regions 3 and 6 in Fig.\,\ref{regions}, for further details see Sect.\,\ref{disc_perturbations}.\\

\noindent  The velocity dispersion decrease at \textit{r}\,$\leq$\,1.5\,arcsec (seen in both tests with \textsc{kinemetry}, Sect.\,\ref{narrow}) could indicate the presence of a $\sigma$-drop. This feature is related to dynamically cold gas (having relatively low $\sigma$) funneled from the outer regions to the nucleus  by a bar during a fast  episode of central gas accretion. From this gas a young stellar population is born with similar velocity dispersion as the accreted gas. This scenario, theoretical modeled by \citet{Wozniak2003}, is a rare phenomenon mostly observed for the stellar kinematics of few nearby spirals (e.g. \citealt{Emsellem2001, Marquez2003, Comeron2008, daSilva2018}). We propose that the $\sigma$-decrease of  $\sim$\,25\,km\,s$^{-1}$ in the inner \textit{r}\,$\leq$\,1.5\,arcsec of NGC\,7469 could be associated to the $\sigma$-drop phenomenon, probing the low-dispersion gas funneled by the bar. So far, in literature there are no previous cases of  $\sigma$-drop observed for the ionised gas kinematics. This detection has been possible only thanks to the exquisite spectral resolution of MEGARA.  To confirm this hypothesis, we search for previous estimates of the stellar kinematics with special emphasis on velocity dispersion radial profiles. We find that the unique measurements  of  the velocity dispersion  of  stars  for NGC\,7469 is by \citet{Onken2004}. They measured the kinematics using the Ca\,triplet at 8498, 8542, and 8662 \AA  \ from long-slit spectroscopic  low-resolution data from Kitt Peak National Observatory  and MDM Observatory (60 and 90 km\,s$^{-1}$  resolution, respectively). The velocity dispersion value is 152\,$\pm$\,16\,km\,s$^{-1}$, which is larger with respect to all the measurements for both narrow components. Their data do not either confirm or discard the $\sigma$-drop scenario. To this aim, a spatially resolved study of the stellar kinematics at high resolution  is required. 

\subsubsection{Detailed comparison between the results from present MEGARA IFS data and previous works}
\label{kin_prev} 
\begin{figure*}
\includegraphics[trim = 2.1cm 13cm 2.2cm 3.1cm, clip=true, width=1.\textwidth]{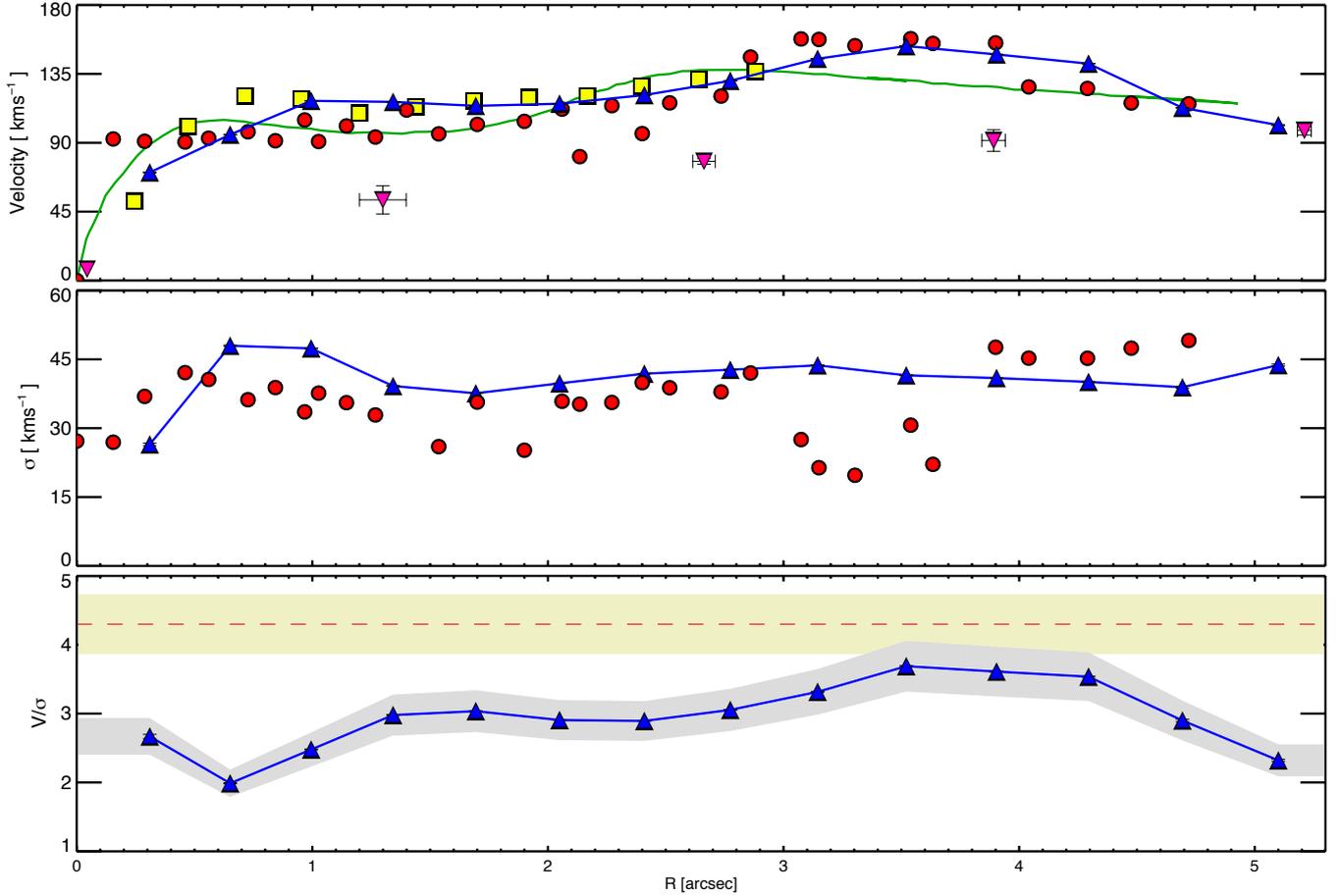}\\
\caption{Rotation curve, velocity dispersion profile and dynamical ratio distribution  as a function of the distance from the photometric center for the narrow component. In all panels, blue triangles mark the values obtained from our kinemetric analysis (Sect.\,\ref{disc_kinemetry}). In the top panel, red circles indicate the rotation curve in MEGARA datacube calculated from the symmetrized rotation curve in the bottom right panel of Fig.\,\ref{pv_sigma} (see also  Sect.\,\ref{PV_diagrams}). Other symbols are from previous works (adapted for comparison purposes). Specifically, magenta upside down triangles indicate the H$\alpha$ rotation curve extracted from the PV-curve presented by \citet{Marquez1994} (their Fig.\,12a); green continuos line corresponds to the model to the CO inclination-corrected rotation curve by \citet{Davies2004} and yellow squares represent the best-fitting to the rotation curve of the HCN emission measured by  \citet{Fathi2015}. Rotation curve by  \citet{MullerSanchez2011} and \citet{Hicks2009} cover only the innermost 1.5\,arcsec, thus are not considered in this figure. In the middle panel, the velocity dispersion radial profiles are from MEGARA data (red circles, Sect.\,\ref{PV_diagrams}) and from the \textsc{kinemetry} modelling (blue triangles, see also Fig.\,\ref{kinemetry_narrow_moments}). In the bottom panel,  the dashed line indicates the global dynamical ratio (V\,/\,$\sigma$\,=\,4.3) estimated in Sect.\,\ref{disc_support_hz}. The shaded yellow and grey bands indicate the positions for which the dynamical ratio is equal to the $\pm$10\,per cent of its value.}      
\label{rVrot} 	
\end{figure*}

\noindent  For the ionised gas,  \citet{AlonsoHerrero2009} modeled the H$\alpha$-[N\,{\small II}] lines using a broad and a narrow component in IFS observations obtained with PMAS in the LARR  mode (1\,arcsec magnification,  V300 grating with a  reciprocal linear dispersion of 1.67\,\AA \ pixel$^{-1}$ and R\,$\sim$\,1000) over a 16\,arcsec\,$\times$\,16\,arcsec field of view. They report a velocity curves ranging from  $-$\,110 to $+$\,207\,km\,s$^{-1}$ (hence $\Delta$V is $\sim$\,159\,km\,s$^{-1}$). This estimate is in  agreement, within uncertainties, with our measurement for the narrow component. The ionised gas velocity dispersion map by \citet{AlonsoHerrero2009} strongly departs from the description of a rotating disc being chaotic. Their measurement of $\sigma_{\rm\,c}$ (125\,$\pm$\,16\,km\,s$^{-1}$) is very different with respect to our estimate for the narrow component but  similar to that of the second component (Table\,\ref{T_properties}).\\
\noindent  The shape of PV-curve of the second component is rather similar to that in  \citet{Marquez1994} obtained from slit spectroscopy of the H$\alpha$-[N\,{\small II}] emission.  The corresponding rotation curve, adapted from their Fig.\,12a, is shown in Fig.\,\ref{rVrot} (magenta triangles). From the same figure we also estimated the peak-to-peak velocity amplitude being of $\sim$\,113\,km\,s$^{-1}$. This estimate is lower than our measurements for both narrow and second components  (Table\,\ref{T_kin}).  This difference could be partially due some perturbations of the rotation curve, especially  relevant at \textit{r}\,$\geq$\,7\,arcsec in the approaching side of the rotating disc (negative velocities).  \\
\noindent  At near-IR wavelengths, the ionised gas kinematics have been derived from the analysis of the Br$\gamma$$\lambda$2.165$\mu$m emission line with SINFONI/VLT and OSIRIS/Keck IFS data by \citet{MullerSanchez2011} in the innermost region (less than 1\,arcsec\,$\times$\,1\,arcsec) at spatial scales of $\sim$\,0.04\,arcsec ($\sim$15\,pc). These authors fitted the observed Br$\gamma$ line profile with a single component finding that the ionised gas velocity dispersion is on average 90\,$\pm$\,9\,km\,s$^{-1}$ (150\,km\,s$^{-1}$ at  maximum) and its velocity field is dominated by rotation.  The same data set has been used by \citet{Hicks2009} with the aim of studying the spatial distribution, kinematics, and column density of the hot molecular gas traced by the H$_{2}$ emission line (1-0\,S(1) at 2.12$\mu$m) in a sample of nearby Seyferts with special emphasis on the correlation of the molecular gas with star formation and AGN properties. For the case of  NGC\,7469, they found ordered rotational motions (e.g. no warps) typical of disc kinematics and a low velocity dispersion (about 60-90\,km\,s$^{-1}$). The latter varies homogeneously over the field of view being not centrally peaked. These results are similar to those measured by \citet{Fathi2015, Davies2004, Izumi2015} for the cold molecular gas. In particular, for  HCN and CO\,\textit{J}\,(2$\rightarrow$1), the rotation curves flatten at $\sim$\,130\,km\,s$^{-1}$ (\citealt{Fathi2015, Davies2004}, see also green line and yellow squares in Fig.\,\ref{rVrot}) with a typical velocity dispersion of 60 and 22 \,km\,s$^{-1}$, respectively. However, all these findings are not fully comparable with those presented in the present work as they strongly differ in terms of type of spatial scale, field of view and spectral resolution (e.g. a factor 5-6 higher to that of SINFONI and OSIRIS).%

\subsection{Dynamical support and disc height of the narrow components }
\label{disc_support_hz}

The comparison of the dynamical ratio (V\,/\,$\sigma$)  between the velocity amplitude and the disc velocity dispersion for the different components allow us to study the different levels of dynamical support. Specifically, the rotational support can be inferred from observed (i.e. no inclination corrected) velocity-to-velocity dispersion ratio, calculated as the ratio between the amplitude  and the mean velocity dispersion across the disc ($\Delta$V and $\sigma$$_{\rm\,avg}$ respectively, Table\,\ref{T_kin}). We found that the  narrow component has a rotation-dominated kinematics (V\,/\,$\sigma$\,=\,4.3) while the second one has an increasingly larger random-motion component (V\,/\,$\sigma$\,=\,1.3). Hence, the low rotational support of the second component is compensated by the large random motions observed and it is suggestive of a thicker, dynamically hotter disc. \\

\noindent  In Fig.\,\ref{rVrot} (bottom panel), we show the radial profile of the dynamical ratio (i.e. V\,/\,$\sigma$) calculated using the kinemetric coefficients resulting from our disc modelling of the velocity field of the narrow component (Sect.\,\ref{disc_kinemetry} and Fig.\,\ref{kinemetry_narrow_moments}). At all radii, V\,/\,$\sigma$ is larger than 2 (the average is 3) suggesting a constant rotational support. The maximum of the V\,/\,$\sigma$-radial profile is 3.7 being consistent, within 16 per cent, with the global dynamical ratio estimation using MEGARA kinematic maps. This value is reached at $\sim$\,3.5\,arcsec, the distance where the PV-curve of the narrow component starts to flatten (Fig.\,\ref{pv_sigma} left). \\ 

\noindent  Using optical images and IFU (FLAMES/GIRAFFE) observations \citet{Flores2006} defined 3 kinematical classes. Briefly these are: \textit{rotating discs}: are characterised by a rotation in the velocity field that follows the optical major axis and the $\sigma$-map shows a clear peak near the galaxy center;   \textit{perturbed discs}: the axis of rotation in the velocity field follows the optical major axis, and the $\sigma$-map shows a off-centred peak or no clear peak; and  \textit{complex kinematics}: system with both kinematic maps that are discrepant to normal rotation discs, and the velocity field is not aligned with the optical major axis. Accordingly to  this simple scheme, both ionised discs in NGC\,7469 would be ascribed to the  \textit{perturbed discs} class, as the kinematic and photometric  axis are nearly coincident (Sect.\,\ref{comparison_narrow}) but velocity dispersion map are not centrally peaked (Sections  \ref{narrow} and \ref{second}). \\
This classification has been recently used by  \citet{Bellocchi2013} to classify the kinematics of ionised gas discs in U/LIRGs. For LIRGs, as NGC\,7469 (log\,(L$_{\rm\,IR}$/L$_{\odot}$)\,=\,11.7, Table\,\ref{T_properties}), the mean (median) value of the dynamical ratios is 3.4 (3.3). Considering both LIRGs and U/LIRGs classified as \lq rotating disc\rq \  and \lq complex kinematics\rq, the dynamical ratios (V\,/\,$\sigma$) are 4.7 and 3.1, respectively. As for comparison, high values of V\,/\,$\sigma$ ($>$\,3) are typical of local spirals \citep{Arribas2008}, and lower V\,/\,$\sigma$ values are found for more distant Lyman break analogs  (i.e. V\,/\,$\sigma$\,$\leq$\,1.1,  \citealt{Goncalves2010}) or thick neutral gas discs in U/LIRGs (i.e. V\,/\,$\sigma$\,$\leq$\,2.5, \citealt{Cazzoli2016}). \\

\noindent Assuming,  as in \citet{Cazzoli2014},  that the narrow components are distributed in a thin (ionised gas) rotating disc and the velocity dispersion is mainly due to the gravitational potential rather than turbulence, an upper limit to the scale height of the disc (\textit{h}$_{\rm\,z}$) can be derived as \textit{h}$_{\rm\,z}$\,=\,$\sigma$$^{2}$$\times$\,R/\,(V(R))$^{2}$ \citep{Cresci2009}.  Considering the inclination corrected semi-amplitude, V(R)\footnote{The values of V(R) for narrow and second components are 324 and 274 km\,s$^{-1}$, respectively. These correspond to the $\Delta$V (Table\,\ref{T_kin}) corrected by inclination  (\textit{i}, Table\,\ref{T_properties}).},  and the mean velocity dispersion across the disc ($\sigma$$_{\rm\,avg}$, Table\,\ref{T_kin}), heights of 20 and 222 parsecs are obtained for the discs traced by the narrow and second components, respectively. These correspond to a  distance of R\,=\,4.3\,arcsec (i.e. 1.43\,kpc, at the adopted distance, Table\,\ref{T_properties}), which is the maximum radius we mapped in the MEGARA cube (Figures \ref{maps_1c} and \ref{maps_2c}). At this distance,  both PV-curves nearly flatten (Fig.\,\ref{pv_sigma}). \\

\noindent Previous works about ionised gas kinematics  in NGC\,7469 (see Sect.\,\ref{kin_prev}) do not include measurements of both the dynamical ratio and disc height, except the work done by \citet{Hicks2009} for the warm molecular gas traced by the H$_{2}$ emission. Although we do not expect a one-to-one correspondence of the ionised and molecular discs, we found that  the estimate of V\,/\,$\sigma$ for the second component is fairly consistent with the previous measurements  by  \citet{Hicks2009}  (see also Sect.\,\ref{kin_prev}). These authors found a low central V\,/\,$\sigma$ ($\sim$\,0.5) indicating that random motions are prominent in the inner region (\textit{r}\,$\sim$\,35\,pc) of NGC\,7469. At  slightly larger distances, i.e. 40\,$\leq$\,\textit{r}\,$\leq$\,140\,pc,  V\,/\,$\sigma$ seems to be constant ($\sim$\,1.0). Nevertheless, their estimate of the disc height strongly differs with ours, being of the order of 30-40\,pc \citep{Hicks2009}. This discrepancy could be either intrinsic (due to the mapping of two different discs, molecular vs. ionised) or to the different assumptions used. They considered the two cases of a disc of self-gravitating stars and gas where the surface density (derived from their estimate of the dynamical mass) is constant and  an isothermal disc for which  V\,/\,$\sigma$\,$\sim$\,\textit{r}\,/\,\textit{h}$_{\rm\,z}$. The latter correspond to the thick-disc approximation \ used in \citet{Cresci2009} (see also \citealt{Cazzoli2014, Cazzoli2016}). We can consider that  the thick disc approximation is more suitable for inferring \textit{h}$_{\rm\,z}$ for the disc traced by the second component as dynamically hotter (likely thicker). In this scenario, an upper limit to  \textit{h}$_{\rm\,z}$ is hence 564\,pc.\\
The   height  of the  ionised (present work) and warm molecular \citep{Hicks2009} discs are well below the measurements of \textit{h}$_{\rm\,z}$ for the thin stellar disc of our Galaxy. As a reference, the stellar thin (thick)  disc in the Milky Way  has a scale height of $\sim$\,200-300 pc   ($\sim$\,1.4\,kpc)  but  with a (much) lower vertical velocity dispersion $\sim$\,20\,km\,s$^{-1}$ ($\sim$\,40\,km\,s$^{-1}$), as reported by \citet{Glazebrook2013}. We would like to note that the thickness of the very-thin disc in NGC\,7469 is comparable to that of the  \lq nuclear stellar disc\rq \ of the Milky Way which have a  scale height of 45\,$\pm$\,5 pc (see \citealt{Nishiyama2013} and \citealt{Rainer2014} for details).\\

\noindent Considering all this, we have found that the ionised gas in NGC\,7469 is distributed in two discs. The one  probed by the narrow component is a very thin disc mainly supported by rotation, in contrast, that traced by the second component is thicker and also dynamically hotter.\\

\noindent The turbulence, due to the  increasing importance of random motions, may result from rapid accretion, disc instabilities, stellar feedback  or possibly thin-disc heating (\citealt{Elmegreen2017} and \citealt{Yoachim2008} and references therein). An alternative origin for the presence of the thickest ionised disc in NGC\,7469 is extraplanar diffuse ionised gas gravitationally bound and virialised  (e.g. \citealt{Levy2018}). Ancillary data, at similar spectral and spatial resolution,  to study the stellar and cold gas kinematic are required to discriminate between these two scenarios. 

\subsection{Velocity Dispersion anomalies}
\label{disc_perturbations}

\begin{figure*}
\includegraphics[width=1.\textwidth]{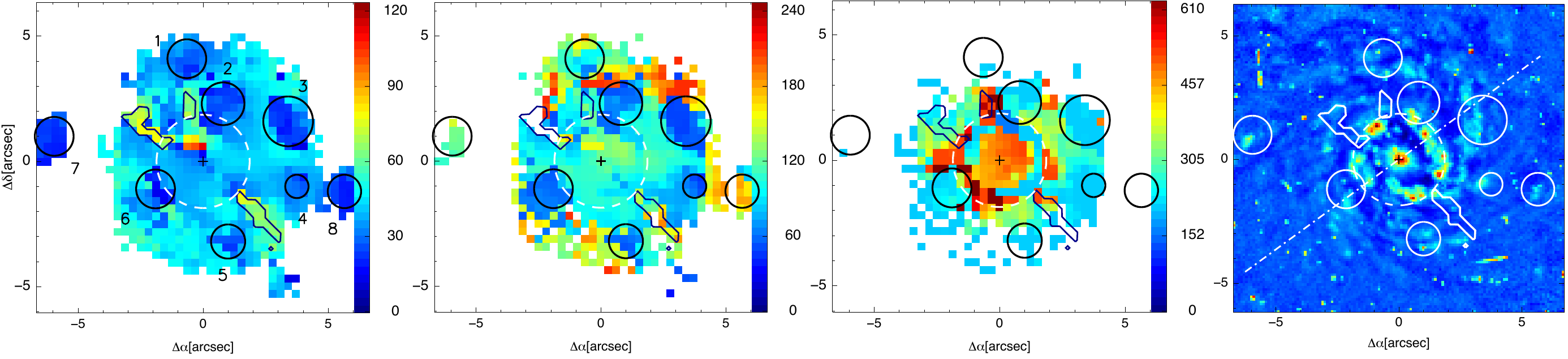}\\
\caption{From \textit{left} to the \textit{right}: the velocity dispersion maps for the narrow, second and intermediate components (see Sect.\,\ref{main_observational_results}) and the \textit{HST} sharp-divided image (see also Fig.\,\ref{HST_morphology}). In all the maps the eight selected regions to be velocity dispersion anomalies (with $\sigma$\,$<$\,60\,km\,s$^{-1}$) are indicated with circles (Sect.\,\ref{disc_perturbations}). These are  labelled (left panel) in a clock-wise pattern and considering their distance to the center (cross). Three iso-$\sigma$ contours mark the anomalies with enhanced velocity dispersion (with $\sigma$\,$>$\,60\,km\,s$^{-1}$) along the minor axis of rotation and outside the nuclear region. See Sect.\,\ref{disc_perturbations} for details about all the $\sigma$-features. The white dashed circle marks the nuclear region, as e.g. in Fig.\,\ref{maps_1c}.  Velocity dispersion maps are in km\,s$^{-1}$ units. The white dot-dashed line marks the photometric major axis as in Fig.\,\ref{maps_1c}.}      
\label{regions} 	
\end{figure*} 


As mentioned in Sections \ref{narrow} and \ref{second}, and \ref{comparison_narrow}, the kinematic maps for the narrow components in Figures \ref{maps_1c} and \ref{maps_2c}  present indications of the presence of  perturbations/deviations of the main disc-like kinematics, especially seen for the velocity dispersion. More specifically, these  \lq velocity dispersion anomalies\rq \ can be of two different kinds. On the one hand,  the velocity dispersion maps of both narrow components reveal a few decrements with clump-like morphology.   On the other hand, a velocity dispersion enhancement along the minor photometric axis  is found only in the $\sigma$-map of the narrow component. All these features are marked in Fig.\,\ref{regions}.

\noindent The first kind of anomalies  correspond to 8 regions (6 in the main disc) where the typical velocity dispersion is lower with respect to those measured either  in the central region or in the main disc (Table\,\ref{T_kin}). More specifically, the velocity dispersions reach values of about $\sigma$\,$\sim$\,20-30\,km\,s$^{-1}$ and $\sim$\,40-60\,km\,s$^{-1}$ for the narrow and second component, respectively. These regions with  low-$\sigma$ marginally overlap with morphological features as clumps of  enhanced H$\alpha$ emission outside the star forming ring  (Figures \ref{HST_morphology} and \ref{regions} right).\\ 

\noindent A similar behavior has been already seen for four galaxies NGC\,2844,  NGC\,4245, NGC\,5953  and NGC\,774 by \citet{FalconBarroso2006}. In their work, where they analyse  the H$\beta$ emission in IFS data from SAURON in its low resolution mode (which delivers a spectral resolution of 4.2\,\AA \ in FWHM) attached to the 4.2-m William Herschel Telescope (WHT), they found low values ($\leq$\,40\,km\,s$^{-1}$) of $\sigma$(H$\beta$)  at the  location of the star-forming ring (bright in H$\beta$). More recently, this behavior has been observed by \citet{Shin2019} in high spatial resolution VLT/MUSE observations of NGC\,5728.\\
\noindent The regions with a velocity dispersion decrement are identified and numbered in the $\sigma$-maps of the narrow component, see  Fig.\,\ref{regions} left (circles). The main selection criteria were: to have low velocity dispersion (i.e. $\sigma$\,$<$\,30\,km\,s$^{-1}$), and to be clearly defined in spatial extension (i.e.  with more than 6 neighboring spaxels). Sizes are between $\sim$\,1\,arcsec and $\sim$\,2\,arcsec (i.e. 327 and 655 pc, respectively, at the adopted distance, Table\,\ref{T_properties}) in diameter. The selection was not intended to match the star-forming clumps clearly visible in \textit{HST} sharp-divided image  (Fig.\,\ref{HST_morphology}), as these are not all clearly visible in the flux maps obtained with MEGARA (Figures \ref{maps_1c} and \ref{maps_2c}, bottom right). \\
\noindent  Table\,\ref{T_pert} reports the average values of velocity dispersion and log\,([N\,{\small II}]/H$\alpha$) for each of the selected regions (circles in Fig.\,\ref{regions}). These measurements are generally homogenous with regions 1, 7, and 8 deviating the most in terms of both velocity dispersion and line ratio. Regions 7 and 8,  located outside the main galaxy disc, seem to have a counterpart in the \textit{HST} image  (Fig.\,\ref{regions}) while it is not the case for region 1. If these three regions are excluded, average values for velocity dispersion and log\,([N\,{\small II}]/H$\alpha$) line ratios are: 24 (55) and $-$\,0.32 ($-$\,0.37) for narrow (second) component, respectively.  These typical values of velocity dispersions for both components at the location of each kinematic perturbation are in agreement with those obtained in previous works  by studying circumnuclear star-forming regions. 
For example, in the case of NGC\,3351, \citet{Hagele2007} found that the H$\beta$ line profiles observed in slit-spectroscopic data from the Intermediate dispersion Spectrograph and Imaging System (ISIS)  on the WHT are well described by two kinematic components with different velocity dispersions, specifically 16\,-\,30\,km\,s$^{-1}$ and 43\,-\,65\,km\,s$^{-1}$. Similar velocity dispersion values are observed in the circumnuclear star-forming regions of other galaxies, e.g. NGC\,7479 and NGC\,6070 (\citealt{Firpo2010}, $\sigma$\,$\sim$\,34\,-\,65\,km\,s$^{-1}$), as well  in samples of either dwarf  (\citealt{Moiseev2012},  $\sigma$\,$\leq$\,40\,km\,s$^{-1}$ generally)  or interacting and isolated galaxies (\citealt{ZaragozaCardiel2015},  $\sigma$\,$\leq$\,60\,km\,s$^{-1}$ typically).  \\
Except two cases (regions 2 and 8 for narrow and second components, respectively), the log([N\,{\small II}]/H$\alpha$) line ratio is generally well below -0.2.
This value could be considered a reference for discriminating between ionisation from AGN and from star formation (\citealt{Kewley2006, Kauffmann2003}; see Sect.\,\ref{ion_mechanisms}).\\
\noindent  All together these measurements seem to suggest suggest that 5 out of 8 regions (i.e. regions from 2 to 6, Fig.\,\ref{regions}) could be associated to the star-forming clumps observed in NGC\,7469 but unresolved in our MEGARA observations. These clumps are producing some perturbations of the overall velocity dispersion maps, but not in the velocity field. This could indicate either a high level of turbulence in these regions or that these are not  fully coplanar with the thin rotating disc of NGC\,7469. \noindent   Nevertheless, we recall that there is not a one-to-one correspondence between the perturbed regions identified in the MEGARA-maps and star-forming regions visible in  the \textit{HST} image (Fig.\,\ref{regions}).  \\

\noindent   For the second kind  of anomalies, excluding the nuclear region, velocity dispersions are in the range between  60 to 80\,km\,s$^{-1}$ along the minor photometric axis (iso-$\sigma$ contours in Fig.\,\ref{regions}). At these locations, log\,([N\,{\small II}]/H$\alpha$) varies between $-$0.39 and $-$0.15 ($-$\,0.27, on average; standard deviation is 0.07 dex). Given the observed line widths and line ratios,  we suggest  that these anomalies are related to either star-formation or mild-shocks (e.g. \citealt{Cazzoli2018}). However, the lack of counterpart in the  \textit{HST}-image makes their origin intriguing being, in any case, difficult to pinpoint with the present data set alone.


\begin{table}
\caption{Average values of velocity dispersion and flux ratio measured at kinematic perturbed regions} 
\begin{tabular}{c c  c c c}
\hline
Component & \multicolumn{2}{c}{Narrow} &\multicolumn{2}{c}{Second} \\
\hline
Region & $\sigma$ & log\,([N\,{\small II}]/H$\alpha$) &  $\sigma$ & log\,([N\,{\small II}]/H$\alpha$) \\
 & km\,s$^{-1}$ & & km\,s$^{-1}$ & \\
\hline
1 & 26 (3) & $-$\,0.40 (0.04) & 87 (15) & $-$\,0.28 (0.06) \\ 
2 & 23 (3) & $-$\,0.16 (0.05) & 58 (5) & $-$\,0.34 (0.03) \\ 
3 & 22 (4) & $-$\,0.36 (0.06) & 51 (5) & $-$\,0.36 (0.03) \\
4 & 27 (4) & $-$\,0.33 (0.07) & 53 (9) & $-$\,0.44 (0.03) \\
5 & 26 (3) &  $-$\,0.39 (0.05) & 57 (3) & $-$\,0.39 (0.01) \\
6 & 20 (6) &  $-$\,0.35 (0.63) & 57 (3) & $-$\,0.32 (0.03) \\
7$^{\dagger}$ & 22 (3) & $-$\,0.44 (0.03) & 116 (10) & $-$\,0.26 (0.03) \\
8$^{\dagger}$ & 22 (4) & $-$\,0.44 (0.03) & 143 (40) & $-$\,0.18 (0.004) \\
\hline
\end{tabular}
\label{T_pert}
\begin{flushleft}
{\textit{Notes.} Regions are marked and numbered in Fig.\,\ref{regions} left; $^{\dagger}$ mark those regions outside the main galaxy disc. In parenthesis the standard deviation.}
\end{flushleft}
\end{table}

\subsection{On the origin of the turbulent emission traced by the intermediate component} 
\label{shock_bar_lense}

As mentioned in Sect.\,\ref{intermediate}, the kinematic maps of the ionised gas, probed by the intermediate component, lack of any rotating-disc feature, being somewhat irregular and chaotic without any peculiar morphology (Fig.\,\ref{maps_3c}, top). We interpret the overall kinematics in these maps as suggestive of the presence of turbulent noncircular motions. Likely explanations for the origin of these motions are twofold. \\
On the one hand, these maps lack  the characteristic  features of outflows detected in starbursts and LIRGs with IFS (e.g. a broad and blue shifted component along the minor axis, see \citealt{Cazzoli2014, Cazzoli2016}). Besides that, the presence of an ionised gas outflow cannot be ruled out. Indeed, recent observations of AGN-driven outflows show that these flows are detected as a broad and blue shifted component but could  not be oriented perpendicular to the galaxy disc, as in the case of NGC\,1068 (\citealt{Garciaburillo2014} and references therein).  Moreover, in NGC\,7469, evidences supporting the presence of an outflow have been found at other wavelengths in two previous works. The wide angle nuclear outflow detected in coronal gas  by  \citet{MullerSanchez2011} has a  similar kinematics with respect to the putative one in MEGARA data, with velocity dispersion up to 250\,km\,s$^{-1}$ and (maximum) velocities $\sim$\,200\,km\,s$^{-1}$. At UV and X-rays multiple blue shifted outflow components have been found but with much larger velocity ($\mid$V$\mid$\,$\geq$\,580\,km\,s$^{-1}$, \citealt{Blustin2007} and reference therein) than those measured in this work (Fig.\,\ref{maps_3c} top). \\
On the other hand, the velocity and velocity dispersion maps  reveal a broad ($\sigma$\,$>$\,450\,km\,s$^{-1}$) and blue shifted emission  ($\mid$V$\mid$\,$<$\,200\,km\,s$^{-1}$) at a distance of 1.7\,arcsec and extending up to  2.9\,arcsec (i.e. from 580 to 990 pc at the adopted distance, Table\,\ref{T_properties}). This region  has an irregular  and patchy ring-like morphology which is unrelated neither to that of the optical continuum or that of the flux emission of the two narrow components. The contours in the \textit{HST} image in  Fig.\,\ref{regions}  (right) highlight that the region with high velocity dispersion is located at the outer edge of the star forming ring and it is unrelated to any  bright clumps of star formation. A possible interpretation is that this turbulent gas is related to the chaotic and turbulent motions associated to gas flows at the Inner Lindblad Resonance (ILR) radius\footnote{The location of the ILR has be proposed to be likely coincident with the ring (with diameter of 3\,arcsec) considering the morphology of the molecular gas,  large-scale bar, optical and radio continuum emission, and  MIR-emission of young stars by  \citet{Wilson1991}, \citet{Marquez1994}, \citet{Davies2004} and \citet{DiazSantos2007}.} of the primary large-scale lens. However, this signature could be rather faint  in inclined galaxies (i.e. not edge-on), and hence  difficult to detect, even more if a bright star-forming ring is present as in the case of NGC\,7469. \\

\noindent  We note that the most redshifted emission in the velocity field of the intermediate component (V\,$>$\,40\,km\,s$^{-1}$) spatially overlaps with 3 regions (numbered as 2, 3 and 4, Fig.\,\ref{regions}) out of the 8 regions analyzed and discussed in Sect.\,\ref{disc_perturbations}.  At these locations the velocity dispersion reaches $\sim$\,200\,km\,s$^{-1}$. This could be possibly explained as due to some turbulent motions outside the plane of the disc associated to these perturbations. \\

\noindent Leaving aside the already discussed features, a possible  interpretation for the remaining emission (with $\sigma$\,$>$\,200\,km\,s$^{-1}$) is that it could probe some diffuse emission  (possibly gravitationally bound to the host galaxy, but not virialized) outside the plane of the discs.


\subsection{Line-ratios and ionisation mechanisms}
\label{ion_mechanisms}
Standard \lq BPT diagrams\rq\ \citep{Baldwin1981} are empirically derived diagrams based on optical emission line ratios (selected to be essentially unaffected by reddening) that allow to discriminate different ionising mechanisms. We cannot use the BPT diagrams to study in detail  the possible ionisation mechanism of the narrow and intermediate components found in our analysis, as the MEGARA VPH (665-HR, see Sect.\,\ref{observations_datared}), at the redshift of NGC\,7469 (Table\,\ref{T_properties}), covers only the H$\alpha$-[N\,{\small II}] complex. Despite this, the values of  log\,([N\,{\small II}]$\lambda$6583/H$\alpha$)  give helpful limits in order to investigate the dominant mechanism of ionisation. \\
 For the following discussion, we do not exclude the regions named \lq disc-perturbations\rq \ (see Sect.\,\ref{disc_perturbations}) as they are relatively small (the diameters are on average of  1.5\,arcsec,  497\,pc at the adopted distance, Table\,\ref{T_properties}) and hence represent a minor fraction of the spaxels within the maps. \\
 Considering the typical values of the histograms for the  [N\,{\small II}]/H$\alpha$ ratio  (Fig.\,\ref{bpts_isto}, averages values are between $-$\,0.35 and $-$\,0.10) and the dividing lines discriminating ionisation from AGN and star formation by \citet{Kewley2006} and \citet{Kauffmann2003}, for all the three components line ratios suggest the ionisation from star-formation as the dominant mechanism.  We do not have spatially resolved information about the [O\,{\small III}]$\lambda$5007/H$\beta$ ratio. However, as the division between AGN and star formation occurs at [O\,{\small III}]/H$\beta$\,$\sim$\,3 (0.5 in log units), such high values of [O\,{\small III}]/H$\beta$, along with low levels of [N\,{\small II}]/H$\alpha$, could have been produced only in gas with very low metallicity. This is unlikely to be the case of NGC\,7469, as the high current star-formation activity\footnote{The ring of star formation, where supernova explosions take place, is contributing up to two thirds of the galaxy's bolometric luminosity (\citealt{Fathi2015} and references therein).} with possible SN-explosions (the last in 2000,  \citealt{Colina2007}) might have considerably polluted with metals the surrounding ISM. \\
For the narrow component the [N\,{\small II}]/H$\alpha$ ratios point to ionisation from star-formation. For  the intermediate component, the distribution of the line ratios outside the nuclear region is skewed towards  Seyfert- and LINER- like ratios (with log\,[N\,{\small II}]$\lambda$6583/H$\alpha$ up to 0.2 dex, Fig.\,\ref{bpts_isto} right). This might suggest  that, given the large velocity dispersion (up to 610\,km\,s$^{-1}$, on average 276\,$\pm$\,8\,km\,s$^{-1}$; see Table\,\ref{T_kin}), shocks play a significant role in the ionisation of the gas \citep{Molina2018} and might have altered the two discs (injecting some level of turbulence).  In the specific case of the second component, although the dominant ionization mechanism is star-formation, with log\,([N\,{\small II}]/H$\alpha$)\,$<$\,$-$\,0.2 generally, for a minor fraction of the data points (35\,per cent) the ratios are between $-$\,0.2 and 0.2 dex suggesting a mixture of star formation, shock excitation and AGN activity. In these particular cases, the knowledge of [O\,{\small III}]/H$\beta$ is essential to pin-point the ionisation mechanism.\\

\noindent The values of the [N\,{\small II}]/H$\alpha$ ratio are in partial agreement with those reported by \citet{AlonsoHerrero2009} for the integrated spectrum (i.e. 0.55, log units), with  the nuclear spectrum as a complete different value (i.e. $-$\,0.26, log units). The values of the [O\,{\small III}]/H$\beta$ ratio  reported in by  \citet{AlonsoHerrero2009} are  0.55 to 1 (in log units) for integrated and nuclear spectra, respectively. Although relatively high values of [O\,{\small III}]/H$\beta$ are already observed, these cannot be used to better constraint the ionisation mechanism of NGC\,7469 due to the difference with our data (e.g. spectral and spatial resolution) and number of components  for the modelling of narrow emission lines with respect to theirs (single vs. multiple Gaussians). 
\noindent Therefore, our measurements of the [N\,{\small II}]/H$\alpha$ ratio indicate the star-formation as the exclusive (dominant) ionisation mechanism of gas probed by the narrow (second) component in NGC\,7469. For the third intermediate broader component, given the observed line widths,  line ratios suggest ionisation by shocks. \\

\noindent Optical nebular line ratios are widely exploited to constrain the metallicity in galaxies (see \citealt{MaiolinoMannucci2019} for a review). Among them,  the only calibrator that can be used with the present MEGARA data is N2 defined to be equal to the log\,([N\,{\small II}]/H$\alpha$). Overall, for the three components, considering the typical line ratios  summarized in column\,9 of Table\,\ref{T_kin}  (outside the nuclear region in order to avoid possible AGN-contamination) and assuming the calibration by \citet{Marino2013}, the oxygen abundance  (12\,+\,log(O/H)) varies from  8.58 to 8.70. These values suggest that the typical  metallicity of the three components is solar (i.e. 8.69,  \citealt{Asplund2009}).


\section[]{Conclusions}
\label{conclusions}

\noindent On the basis of optical MEGARA IFS high resolution (R$\sim$\,20\,000) data we have studied the  properties (kinematic and dynamical as well as fluxes-ratios and oxygen abundances) of the ionised gas in the Seyfert\,1.5 galaxy NGC\,7469, using as tracers the H$\alpha$-[N\,{\small II}] emission lines. \\
The conclusions of the present study can be summarised as follows:
\begin{enumerate}
\item[(1)] \textit{BLR properties}. In the  nuclear region of NGC\,7469 (\textit{r}\,$\leq$\,1.85 arcsec)  the  broad ($\sigma$\,=\,1100\,$\pm$\,10\,km\,s$^{-1}$, FWHM\,$\sim$\,2590\,km\,s$^{-1}$)  H$\alpha$ component is dominating (i.e contribution of 41 per cent)  the global H$\alpha$-[N\,{\small II}] profile,  being originated in the (unresolved) BLR of the AGN.  The unprecedented spectral resolution of MEGARA allows a multi-component modelling of the H$\alpha$-[N\,{\small II}] blend (not accounted for in previous works), that results in the most accurate measurements so far for the BLR-originated  H$\alpha$ component.
\item[(2)] \textit{Discs kinematics and classification}. 
The two discs, probed by the narrow and second components,  nearly co-rotate with similar peak-to-peak  velocities, 163 and 137\,km\,s$^{-1}$, respectively but with different velocity dispersion, i.e. 38\,$\pm$\,1 and 108\,$\pm$\,4\,km\,s$^{-1}$, respectively. The analysis of their kinematic maps (velocity and velocity dispersion)  led to be both classified as \lq Perturbed discs\rq \ \citep{Flores2006} since their  major kinematic axis  well aligned with the photometric axis, their velocity maps are fairly regular but both velocity dispersion maps deviate from the case of an ideal rotating disc. Although both components share the same disc-classification, we remark that the disc traced by the second component is the most perturbed within the two, with a more disturbed velocity field and irregular PV-curve.  

\item[(3)] \textit{Dynamical support and disc height}. The difference in the velocity dispersion of the two discs  indicates that the disc traced by the narrow component share a similar dynamical status with local spirals, while the other, traced by the second component, is likely to be  thicker and turbulent, having higher velocity dispersion with respect to both spirals and high-z star forming galaxies. We found that the very thin (20\,pc) ionised gas disc, mainly supported by rotation (V/$\sigma$\,=\,4.3), is embedded in a thicker (222\,-564\,pc), dynamically hotter  (V/$\sigma$\,=\,1.3) one. 

\item[(4)] \textit{Discs modelling and kinemetric analysis}. We successfully modeled  the kinematics of the thin (\textit{h}$_{\rm\,z}$\,=\,20\,pc) ionized gas disc with \textsc{kinemetry} \citep{Krajnovic2006}. The position angle is remarkably stable (within 120$^{\circ}$ and 130$^{\circ}$) at nearly all reconstructed ellipses, with no indication of kinematic twists, counter rotating cores or kinematically distinct cores.  These values are in good agreement with literature (126$^{\circ}$) and other estimates in the present work, i.e. (125\,$\pm$\,10)$^{\circ}$ from MEGARA maps and  (124.3\,$\pm$\,2.5)$^{\circ}$ from an alternative method by \citet{Krajnovic2006}. In addition, it exhibits a kinematically round velocity map with large opening angles reflected in the high values of (\textit{q}), i.e. 0.88 on average, and with intermediate properties between \textit{regular} and \textit{non-regular rotators} \citep{Krajnovic2011} as indicate by the \textit{k}$_{5}$/\textit{k}$_{1}$ parameter ($\sim$\,0.04).

\item[(5)] \textit{Ionised gas $\sigma$-drop}. We found a velocity dispersion diminution in the velocity dispersion radial profile (output of \textsc{kinemetry}), of about 25\,km\,s$^{-1}$ at \textit{r}\,$\leq$\,1.5\,arcsec. This feature is suggestive of the presence of a $\sigma$-drop related to dynamically cold gas funneled from the outer regions to the nucleus  by a bar during a fast  episode of central gas accretion. So far, in the literature there are no previous cases of  $\sigma$-drop observed for the ionised gas kinematics (with few cases detected through the analysis of stellar kinematics). This detection has been possible only thanks to the exquisite spectral resolution of MEGARA.

\item[(6)] \textit{Intermediate-width component}. The morphology  and the kinematics of the third (intermediate-width) component is suggestive of the presence of turbulent noncircular motions, possibly associated either to an ionised gas wide angle outflow (oriented not perpendicular to the galaxy disc) or to gas flows related to the large-scale lens. Part of the ionised gas traced by this component could be also due to turbulent motions outside the plane of the disc related to disc perturbations, and with diffuse  ionised gas gravitationally bound to the host galaxy, but not virialized.

\item[(7)] \textit{Ionisation mechanisms and Oxygen abundances}.  For the narrow (second) component the [N\,{\small II}]/H$\alpha$ line ratios are indicative of  star-formation as the unique (dominant) mechanism of ionisation. For the intermediate component, given the observed line-widths,  the [N\,{\small II}]/H$\alpha$ ratios are consistent with  ionization from shocks. For all the three kinematic components our measurements suggest that the gas has roughly solar metallicity.

\end{enumerate}

\noindent As a final remark, we highlight that studies of this kind of nearby galaxies at high spectral resolution represent a benchmark for the interpretation of future observations with next generation 30m-class telescopes. For example, the European Extremely Large Telescope,  will be equipped with intermediate  to high resolution (R$\sim$\,3\,000\,-\,20\,000) spectrographs such as MOSAIC and HARMONI.


\section*{Acknowledgements}

This paper is based on observations made with the Gran Telescopio Canarias (GTC), installed in the Spanish Observatorio del Roque de los Muchachos of the Instituto de Astrof{\'i}sica de Canarias, in the island of La Palma. This work is based on data obtained with MEGARA instrument, funded by European Regional Development Funds (ERDF), trough Programa Operativo Canarias FEDER 2014-2020.\\
SC, IM, JM and LHM acknowledge financial support by the Spanish Ministry of Economy and Competitiveness (MEC) under grant no. AYA2016-76682-C3. We also acknowledge financial support from the State Agency for Research of the Spanish MCIU through the \lq\lq Center of Excellence Severo Ochoa\rq\rq \ award to the Instituto de Astrof{\'i}sica de Andaluc{\'i}a (SEV-2017-0709).\\
AGdP, JG, ACM, SP, and NC  acknowledge financial support from the Spanish  MEC under grant no. AYA2016-75808-R.\\
This research has made use of the NASA/IPAC Extragalactic Database (NED), which is operated by the Jet Propulsion Laboratory, California Institute of Technology, under contract with the National Aeronautics and Space Administration.
We acknowledge the usage of the HyperLeda database (\url{http://leda.univ-lyon1.fr}).\\
This paper made use of the plotting package \textsc{jmaplot}, developed by Jes{\'u}s Ma{\'i}z-Apell{\'a}niz  available at: \url{http://jmaiz.iaa.es/software/jmaplot/ current/html/jmaplot_overview.html}. \\
The authors acknowledge  the anonymous referee for her/his instructive comments that helped to improve the presentation of this paper.\\
SC thanks R.\,Amorin and R.\,Sch{\"o}del for their useful comments.
\bibliographystyle{mn2e}
\bibliography{Bibliography.bib}


%


\label{lastpage}
\end{document}